\documentclass[twocolumn,pra,aps,reprint,superscriptaddress]{revtex4}
\usepackage{graphicx}
\usepackage{dcolumn}
\usepackage{bm}
\usepackage{bbm}
\usepackage[utf8]{inputenc}
\usepackage{amssymb}
\usepackage{amsmath}
\usepackage{bbold}
\usepackage{pstricks}
\usepackage{color}
\usepackage{slashed}
\usepackage{braket}
\usepackage{natbib}
\usepackage{float}
\usepackage{verbatim}
\usepackage{feynmp-auto}
\usepackage{siunitx}
\usepackage{lipsum}

\newcommand{\Tr}{\operatorname{Tr}}

\graphicspath{ {./figures/} }
\begin{document}

\title{Inflationary preheating dynamics with ultracold atoms}

\author{T. V. Zache}
 \email[]{zache@thphys.uni-heidelberg.de}
\affiliation{Institut f\"{u}r Theoretische Physik, Universit\"{a}t Heidelberg, Philosophenweg 16,
69120 Heidelberg, Germany}

\author{V. Kasper}
\affiliation{Physics Department, Harvard University, 17 Oxford Street, Cambridge MA 02138, USA}

\author{J. Berges}
\affiliation{Institut f\"{u}r Theoretische Physik, Universit\"{a}t Heidelberg, Philosophenweg 16,
69120 Heidelberg, Germany}

\date{\today}

\begin{abstract}
We discuss the amplification of loop corrections in quantum many-body systems through dynamical instabilities.
As an example, we investigate both analytically and numerically a two-component ultracold atom system in one spatial dimension. 
The model features a tachyonic instability, which incorporates characteristic aspects of the mechanisms for particle production in early-universe inflaton models. We establish a direct correspondence between measureable macroscopic growth rates for occupation numbers of the ultracold Bose gas and the underlying microscopic processes in terms of Feynman loop diagrams. We analyze several existing ultracold atom setups featuring dynamical instabilities and propose optimized protocols for their experimental realization. 
We demonstrate that relevant dynamical processes can be enhanced using a seeding procedure for unstable modes and clarify the role of initial quantum fluctuations and
the generation of a non-linear secondary stage for the amplification of modes.
\end{abstract}


\maketitle

\section{Introduction and overview}

Ultracold quantum gases provide a unique opportunity for studying nonequilibrium phenomena that are otherwise very difficult to access experimentally. Ensembles of trapped atoms can be largely isolated from the environment, such that they offer the possibility to address fundamental aspects of quantum many-body systems governed by unitary time evolution. In addition, cold-atom setups provide a very flexible testbed, with tunable interactions or with reduced dimensionality by shaping the confining optical potential, realizing a multitude of different relevant physical situations.

It has been pointed out that characteristic aspects of the  evolution of the early universe may be addressed with table-top experiments~\cite{schmiedmayer2013cold}, such as acoustic oscillations in the time evolution of an ultracold quantum degenerate gas~\cite{hung2013cosmology}, or inflationary quasiparticle creation and thermalization dynamics in coupled Bose-Einstein condensates~\cite{opanchuk,posazhennikova2016inflationary}. The latter may involve paradigmatic dynamical phenomena such as nonequilibrium instabilities seeded by quantum fluctuations~\cite{Felder:2001kt}, which trigger important nonlinear phenomena such as enhanced particle creation from secondary instabilities~\cite{berges2003parametric} and the subsequent approach to nonthermal fixed points with Bose condensation~\cite{Berges:2008wm} long before thermalization sets in.  

In this work we analyze how two-component bosonic cold-atom systems in one spatial dimension may be used to address relevant aspects of the nonequilibrium dynamics of tachyonic instabilities arising in the context of inflationary models~\cite{Felder:2001kt,Felder:2000hj}. Experimentally, we have in mind the twin-atom beam experiments of the Schmiedmayer laboratories~\cite{bucker2011twin}, and the spin-changing collision dynamics arising in bosonic spin-one systems as employed by the Oberthaler group~\cite{linnemann2016quantum}.
We identify the relevant dynamical processes of the cold-atom system and describe them using quantum-statistical field theory. In this way we establish a one-to-one correspondence between macroscopic observables of the Bose gas and microscopic processes in terms of Feynman loop diagrams for correlation functions. 

\begin{figure*}
	\centering{
		\includegraphics[scale=0.3,angle=-90]{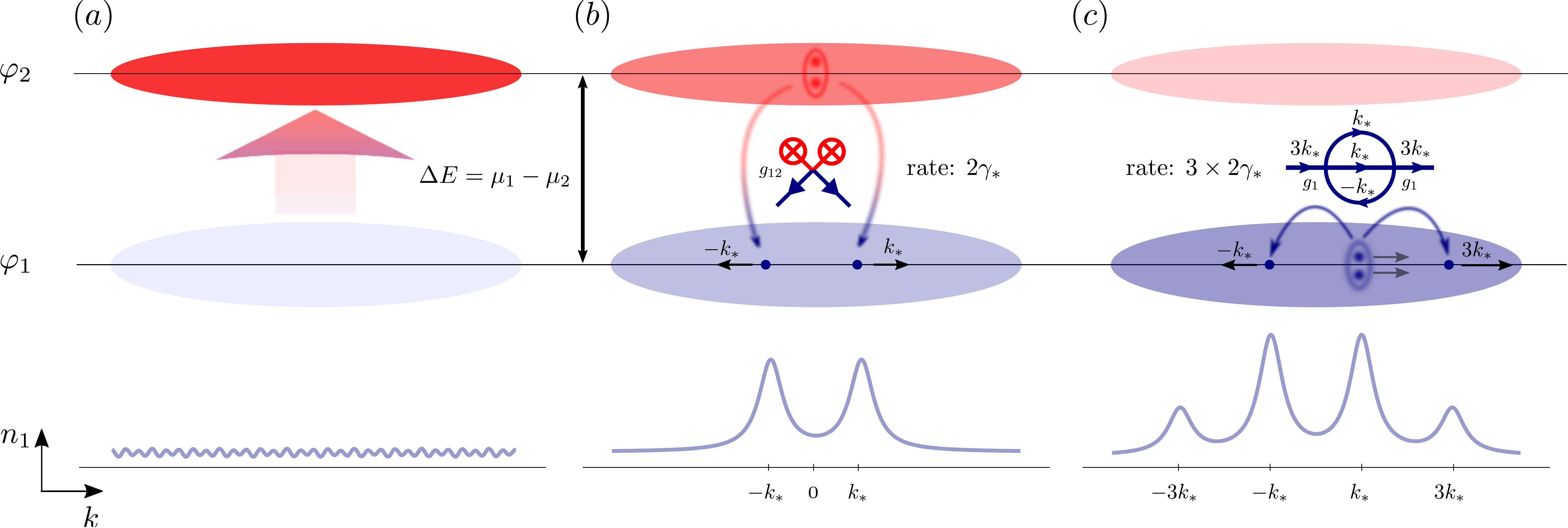}}
	\caption{\label{secondary_pictorial}Generation of primary instabilities and secondaries for the Bose gas with components $\varphi_1$ and $\varphi_2$. $(a)$ Initial condition: The system is prepared in the `excited' state $\varphi_2$ with energy gap $\Delta E$ between the components. $(b)$ Primary instability: The condensate decays with the excitation of unstable modes having momenta $\pm k_*$ and chararacteristic primary growth rate $2 \gamma_*$. $(c)$ Secondaries: Subsequently, nonlinear processes result in the enhanced growth of occupation numbers in a wider momentum range, e.g.\ at $\pm 3 k_*$ as indicated in the figure. The number of internal lines $l$ of the corresponding Feynman loop diagram for the nonlinear process determines the enhanced growth rate as $l \times 2\gamma_*$, e.g.\ $3 \times 2\gamma_*$ for the two-loop diagram shown.}
\end{figure*}

More precisely, we first analyze the generation of instabilities that are seeded from initial quantum fluctuations in the Bose gas. These primary instabilities trigger an exponential increase of the occupation numbers of characteristic unstable modes with momenta $\pm k_*$. This early stage is followed in time by a secondary growth period, which describes the nonlinear enhancement of occupancies in a wider momentum range. In FIG.~\ref{secondary_pictorial} we sketch the initial state and the subsequent processes for the Bose gas with components $\varphi_1$ and $\varphi_2$.
The initially unstable state $\varphi_2$ decays via the excitation of characteristic modes of $\varphi_1$ at momenta $\pm k_*$. As these modes become highly occupied, their interactions lead to the excitation of modes with higher momenta of multiples of $k_*$. Remarkably, macroscopically measurable growth rates and their characteristic momenta give, in principle, direct information about the topology of the Feynman diagrams that describe the underlying microscopic processes.           

While for typical inflaton models the secondaries have a significant impact on the dynamics, their role can be restricted by total particle number limitations in cold-atom experiments. In particular, we demonstrate that for the existing twin-atom beam setup~\cite{bucker2011twin} the experimental detection of the relevant higher-loop processes may be difficult. To overcome this limitation, we propose to seed the primary instabilities: Instead of
preparing the whole condensate in the `excited' state $\varphi_2$,
one keeps a fraction of all atoms in the `lower' state $\varphi_1$ and imprints a modulation to transfer them into the most unstable modes $\pm k_*$. We perform classical-statistical (Truncated Wigner) simulations to numerically support our findings. 

This publication is organized as follows: 
In section \ref{sec:2} we discuss our model system and relate it to possible experimental realizations.
Subsequently, we analytically determine the early-time dynamics of this many-body system including the primary instability in section~\ref{sec:3}.
In the following section~\ref{sec:4} we demonstrate the existence of secondaries being present in this system numerically and give a simple heuristic argument for their generation.
In section \ref{sec:5} we complement the observations with a more refined explanation in terms of Feynman diagrams, which allows us to estimate the secondaries' growth rates.
Furthermore, we propose a seeding strategy to make these non-linear phenomena experimentally better accessible in section \ref{sec:6}. 
To get a better understanding of the situation with and without seed, we consider a classical toy model that encompasses the necessary ingredients for the generation of secondaries in section~\ref{sec:7}. 
Finally, we conclude in section~\ref{sec:8}.

\section{Two-component Bose gas \label{sec:2}}

The inflationary early universe undergoes very rapid expansion, after which all energy ends up in a vacuum-like state
without entropy or particles. Typical models describe this state in terms of a coherent Bose field, the inflaton, whose decay finally leads to the observed particle content of today's universe. Very efficient mechanisms for particle production from inflaton decay involve nonequilibrium instabilities, such as parametric resonance~\cite{kofman1994reheating,kofman1997towards} or tachyonic instabilities~\cite{Felder:2001kt,Felder:2000hj}. The latter can be particularly fast and nonlinear many-body phenomena beyond simple classical-field or Bogoliubov--type approximations play a crucial role. 

A prominent class of inflationary scenarios -- hybrid inflation -- involve multiple scalar fields~\cite{linde2008inflationary}. In this work, we consider a model with a bosonic complex field operator $\varphi_a$ with two components $a=1,2$. This model allows us to address  essential features of instability dynamics and the build-up of non-linearities, where we will concentrate on tachyonic instabilities associated to imaginary parts in the dispersion relation. While the couplings to other degrees of freedom of the Standard Model of particle physics play an important role at later times, we focus on the bosonic self-interactions that dominate the dynamics at earlier times of the nonequilibrium evolution.
It should be emphasized that, even though typical couplings in cosmological inflaton models are usually very small, the dynamics become strongly correlated because of the instabilities that amplify small nonlinear corrections exponentially fast. 

Although the time scales depend on the specifics of any microscopic model chosen, essential aspects of the dynamics can be be studied in a wide range of different many-body systems. This is based on the observation that there is an effective loss of details about the initial conditions and model parameters as time proceeds. In recent years this culminated in the discovery of new universality classes far from equilibrium~\cite{Berges:2008wm}, which characterize a significant part of the systems'
dynamical evolution in terms of scaling exponents and scaling functions that are the same for a wide range of different relativistic models~\cite{Berges:2008wm,Berges:2008sr,Berges:2010ez,Gasenzer:2011by,Berges:2012us,Gasenzer:2013era,orioli2015universal} as well as non-relativistic systems of ultracold quantum gases~\cite{Scheppach:2009wu,Nowak:2010tm,Nowak:2011sk,Berges:2012us,orioli2015universal,Schachner:2016frd,karl2016strongly}. 

In an ultracold atom setup with different field components $\varphi_a$, 
the index $a$ typically labels internal states of the 
bosonic atoms such as hyperfine states or eigenstates within an 
external trapping potential. The quantum many-body Hamiltonian of our model system is given by
\begin{align}
\!\!\!\!\!\!H &= \sum_{a=1}^2 \int dz \left\{ \frac{|\nabla_z \varphi_a|^2}{2M} +\left[V(z)- \mu_a\right]|\varphi_a|^2 + \frac{g_a}{2}|\varphi_a|^4 \right\} \notag \\
&+\frac{g_{12}}{2} \int dz  \left\{ (\varphi^\dagger_1)^2 \varphi^2_2 + 4|\varphi_1|^2 |\varphi_2|^2 + (\varphi^\dagger_2)^2 \varphi^2_1 
\right\} \,  , \label{eq:Hamiltonian1}
\end{align}
where we abbreviated $|\varphi_a|^2 = \varphi^{\dagger}_a \varphi_a $ and $|\varphi_a|^4 = \varphi^{\dagger}_a \varphi^{\dagger}_a \varphi_a \varphi_a $.
The interaction constants $g_a$ and $g_{12}$ are related to 
intra- and interspecies scatterings, respectively, and we only consider the repulsive case $g_a,g_{12} > 0 $. $M,V(z)$ and $\mu_a$ are the atomic mass, the one-dimensional trapping potential and two chemical potentials, respectively. The latter controls the total particle number $N$ and the energy gap $\Delta E = \mu_1 - \mu_2$ between the two components. 

The system with Hamiltonian (\ref{eq:Hamiltonian1}) is invariant under a global $\mathcal{O}(2)$ transformation, $(\varphi_1, \varphi_2) \rightarrow (\varphi_1,\varphi_2)e^{i\theta}$, which implies total particle number conservation. For $g_{12} = 0$, the symmetry group is enlarged to $\mathcal{O}(2) \times \mathcal{O}(2)$, i.e. $(\varphi_1, \varphi_2) \rightarrow (\varphi_1 e^{i\theta_1},\varphi_2 e^{i\theta_2})$ corresponding to individual number conservation. For the special case of $g_1 = g_2 = g_{12}$, the interaction part is invariant under $\mathcal{O}(2) \times \mathbb{Z}_2$ due to the symmetry $(\varphi_1, \varphi_2) \rightarrow (\varphi_2, \varphi_1)$. 

The above Hamiltonian can be realized in various ways. We specifically have in mind two different experimental setups that rely on external or internal degrees of freedom and should allow the observation of the dynamical features to be discussed in the following sections. The first experimental setup concerns the twin-atom beam experiment \cite{bucker2011twin}. The authors realized an effectively one-dimensional Bose gas by a magnetic atom-chip trap and applied sophisticated optimal-control techniques to access specific excited states of the radial trapping potential, which in turn serve as the required two components. More precisely, the full three-dimensional system is described by the Hamiltonian
\begin{align}
H_{3D} = \int d^3 r \; \left[\frac{\left| \nabla \psi\left(\mathbf{r}\right)\right|^2}{2M} +  V(\mathbf{r})  \left|\psi\left(\mathbf{r}\right)\right|^2+ \frac{g}{2} \left| \psi\left(\mathbf{r}\right)\right|^4 \right] \; ,
\end{align}
where $V(\mathbf{r})$ is the trapping potential in three spatial dimensions (3D) and $g$ denotes the interaction constant. The fields $\varphi_a(z)$ depending only on the (1D) spatial $z$-direction arise as coefficients in an expansion of the 3D field operator, taken as 
\begin{align}
\psi(\mathbf{r}) = \varphi_1(z) \psi_1(x,y) + \varphi_2(z) \psi_2 (x,y) + \dots \; ,
\end{align}
where $\psi_a(x,y)$ denote the two lowest lying eigenfunctions of $V(\mathbf{r})$. Performing the $x,y$ integrals in the grand-canonical Hamiltonian corresponding to $H_{3D}$ and truncating the above expansion yields a Hamiltonian of the form $\eqref{eq:Hamiltonian1}$. This truncation is expected to be a valid approximation if the radial potential is chosen such that the energetically higher lying states are not excited. The precise form of the radial potential then determines the parameters entering the two-component model in consideration (cf.\ appendix \ref{appendix_parameter}).

As an alternative to exciting states of the external potential, one may utilize internal degrees of freedom of spinor Bose gases \cite{kawaguchi2012spinor}. In such setups, the gas is usually rendered effectively one-dimensional by optical dipole traps. We emphasize that our model system is closely related to spin-1 systems but can not be mapped onto so-called pseudo-spin-1/2 systems. The latter have been extensively studied, e.g.\ as binary mixtures in the context of the miscible-immiscible phase transition~\cite{sabbatini2011phase}. It is necessary to consider larger spins because the Hamiltonian \eqref{eq:Hamiltonian1} differs from a pseudo-spin-1/2 system by the ``spin-changing" terms
\begin{align}
h_\text{SC}\sim \left[ (\varphi^\dagger_1)^2 \varphi^2_2 + \text{h.c.} \right] \; .
\end{align}
These terms turn out to play a crucial role for the dynamical instabilities considered in this work. Experimentally, spin-changing collisions arise naturally in bosonic spin-1 systems and have been utilized, e.g., for the creation of entangled twin-atom states \cite{gross2011atomic}. In such a setup, the bosonic gas is confined to a hyperfine state $F=1$ with three magnetic quantum numbers $m_F= 0, \pm 1$. The spin changing terms are then given by
\begin{align}
\tilde{h}_\text{SC} &\sim \left[ \tilde{\varphi}^\dagger_1 \tilde{\varphi}^\dagger_{-1} \tilde{\varphi}_{0} \tilde{\varphi}_0 + \text{h.c.}\right] \nonumber\\ &=  \frac{1}{2} \left[ \left(\tilde{\varphi}_S^\dagger\right)^2 \tilde{\varphi}_0^2 + \left(\tilde{\varphi}_A^\dagger\right)^2 \tilde{\varphi}_0^2 + \text{h.c.}\right] \; ,
\end{align}
where $\tilde{\varphi}_{m_F}$ label the spinor components and we have introduced $\tilde{\varphi}_S = \frac{1}{\sqrt{2}} \left(\tilde{\varphi}_1 + \tilde{\varphi}_{-1}\right)$ and $\tilde{\varphi}_A = \frac{i}{\sqrt{2}} \left(\tilde{\varphi}_1 - \tilde{\varphi}_{-1}\right)$. Thus $m_F = 0$ corresponds to $\varphi_2$, and $\varphi_1$ would be doubly degenerate as the (anti-) symmetric combinations of $m_F = \pm 1$. In the presence of the quadratic and negligibly small linear Zeeman effect, the resulting Hamiltonian mainly differs from \eqref{eq:Hamiltonian1} by this degeneracy and we restrict ourselves to the model containing only two components for simplicity. Nevertheless, we will argue that our results only depend on a few generic ingredients, such that many aspects also apply to the system with three components. For completeness, we note that the full spin-1 Hamiltonian can be found, e.g., in Ref.~\cite{uchino2010bogoliubov}. 

\section{Primary Instabilities \label{sec:3}}

Before we turn to simulations in section \ref{sec:4}, we first obtain an analytic understanding of primary instabilities. This is done by linearizing the equations of motion in fluctuations around the initial condensate. We identify instabilities as exponentially growing solutions characterized by a dispersion relation with non-vanishing imaginary part.

In recent twin beam experiments \cite{bucker2011twin, wasak2014bogoliubov} it was possible to transform
all atoms from the lower state $\varphi_1$ to the energetically higher state $\varphi_2$. To lowest order in fluctuations, this configuration
is considered to be a stationary state which we denote by $\varphi_{2,s}$. The latter is the solution of the equation
\begin{align}
\mu_2 \varphi_2 = \left[ -\frac{\nabla^2_z}{2M} + V(z) + g_2 |\varphi_2|^2  \right] \varphi_2 \,.
\end{align}
In order to detect the dynamical instability, we describe the evolution of $
\varphi_1$ by using the approximate Hamiltonian 
\begin{align}
H &\simeq \int dz \left[ \frac{|\nabla_z \varphi_1|^2}{2M} +\left(V(z)- \mu_1 + 2 g_{12} |\varphi_{2,s}|^2 \right)|\varphi_1|^2\right] \notag \\
&+\frac{g_{12}}{2} \int dz  \left[ (\varphi^\dagger_1)^2 \varphi^2_{2,s}  + (\varphi^\dagger_{2,s})^2 \varphi^2_1 
\right] \, .
\end{align}
The Heisenberg equations of motion for the atoms then read
\begin{align}
i \partial_t  \varphi_1 &=h_0  \varphi_1 +  h_1 \varphi^{\dagger}_1 \, ,
\end{align}
where we have abbreviated
\begin{align}
h_0 [\varphi_{2,s}] &= \left(  -\frac{\nabla^2_z}{2M} -\mu_1 + V(z) + 2g_{12} |\varphi_{2,s}|^2 \right) \, ,
\\
h_1 [\varphi_{2,s}] &= g_{12}\varphi_{2,s}^2  \, .
\end{align}
The approximate dynamical equations for $\varphi_1$ can be solved by a Bogoliubov transformation
\begin{align}
 \varphi_1(t,z) &= \sum_k \left[ u_k(z) a_k e^{-i\omega_k t } + v_k^*(z) a^{\dagger}_k e^{i\omega_k^* t } \right] \; ,
\end{align}
which leads to the Bogoliubov-de-Gennes equations 
\begin{align}
&\begin{pmatrix}
h_0	 & h_1 \\ -h_1 & -h_0
\end{pmatrix}\begin{pmatrix}
 u_k(z) \\ v_k(z)
\end{pmatrix}\! =\! \omega_k \!
\begin{pmatrix}
 u_k(z) \\ v_k(z)
\end{pmatrix} \, .
\label{eq:BdG}
\end{align}
We refer to the appendix \ref{appendix_bogoliubov} for details of the Bogoliubov 
transformation in the presence of unstable modes. The eigenvalues $\omega_k$ of (\ref{eq:BdG}) and the Bogoliubov mode functions $u_k$ and 
$v_k$ determine the low-energetic excitations and thus the quasi-particle spectrum. We proceed by specifying the trapping potential as either $a)$ harmonic $V(z) = \frac{m}{2}\omega^2 z^2$, or $b)$ a box potential $V(z) = V_\infty \Theta \left(|z| - \frac{L}{2}\right)$ with $V_\infty \rightarrow \infty$. In the former case, we use the stationary solution $\varphi_{2,s}^{(a)}= \sqrt{\frac{\mu_2}{g_2}\left(1- \frac{z^2}{R_{\text{TF}^2}}\right)}$ within the Thomas-Fermi (TF) approximation with the TF radius given by $R_{\text{TF}} = \sqrt{\frac{2 \mu_2}{M w^2}}$. The normalisation of $\varphi_2$ relates the particle number $N_2$ and the chemical potential. For case $b)$, we obtain $\varphi_{2,s}^{(b)} = \sqrt\frac{\mu_2}{g_2}$ with $\mu_2 = n_2 g_2$ where $n_2 = \frac{N_2}{L}$ is the homogeneous particle density. Going from case $a)$ to case $b)$ corresponds to the replacement
\begin{equation}
\label{approx}
\frac{1}{2R_{TF}} \int dz \, e^{iqz} \varphi_{2,s}^{(b)}(z)  \rightarrow \delta(q)
\end{equation}
in the equations of motion in momentum space.
Solving these equations, we arrive at the squared dispersion relation
\begin{align}
\omega_k^2 = \left( \frac{ k^2}{2M} + \frac{2g_{12}\mu_2}{g_2}- \mu_1 \right)^2 - \left( \frac{g_{12}\mu_2}{g_2}\right)^2 \,.
\label{disp_rel}
\end{align}
The system becomes unstable for 
$\omega_k^2 < 0$ with the two most unstable modes being
\begin{align}
\label{primary}
\pm k_* = \pm \sqrt{2M\left(\mu_1 - \mu_2\frac{2 g_{12}}{g_2}\right)}
\end{align}
as determined from the maximal $-\omega_{k_*}^2 \equiv \gamma_*^2 = \left(\frac{g_{12}}{g_2}\mu_2\right)^2$.
Because of the symmetry $k\rightarrow -k$ in the 
Bogoliubov spectrum, one produces two atom 
beams in opposite directions with momentum $k_*$.
Henceforth, these beams are termed twin 
beams. 
The bandwidth of the instability is given by
\begin{align}
\label{fineness}
\delta k = \sqrt{2M}\left(\sqrt{\mu_1 - \mu_2 \frac{g_{12}}{g_2}}-\sqrt{\mu_1 - \mu_2 \frac{3g_{12}}{g_2}}\right)  
\end{align}
Because of the property $\omega_k^2 < 0$ in the instability regime, they are
called tachyonic instabilities in a cosmological context~\cite{Felder:2001kt,Felder:2000hj}.
The initial growth of the occupation of atoms associated with the most 
unstable mode is 
\begin{align}
\frac{1}{2} \braket{\left\lbrace\varphi_1^{\dagger}(t,k_*), \varphi_1(t,k_*)\right\rbrace} \simeq e^{2\gamma_* t} \, ,
\end{align}
where $\varphi(t,k)$ is the momentum mode appearing in the Fourier expansion $\varphi(t,x) = \frac{1}{\sqrt{L}}\sum_k e^{-ikx} \varphi(t,k)$. Note that the above growth rate has been calculated for the homogeneous case $b)$ and serves as a reference value for the trapped case $a)$, where nearby modes will also contribute to the instability (cf. eq. \eqref{approx}). Since these generally grow slower, we expect that the true most unstable growth rate $2\bar{\gamma}_1$ will effectively be damped, i.e.
\begin{align}
\bar{\gamma}_1 \lesssim \gamma_* = \frac{g_{12}}{g_2}\mu_2 \; .
\end{align}

\section{Secondaries \label{sec:4}}

\begin{figure*}
	\centering{
		\includegraphics[scale=0.25]{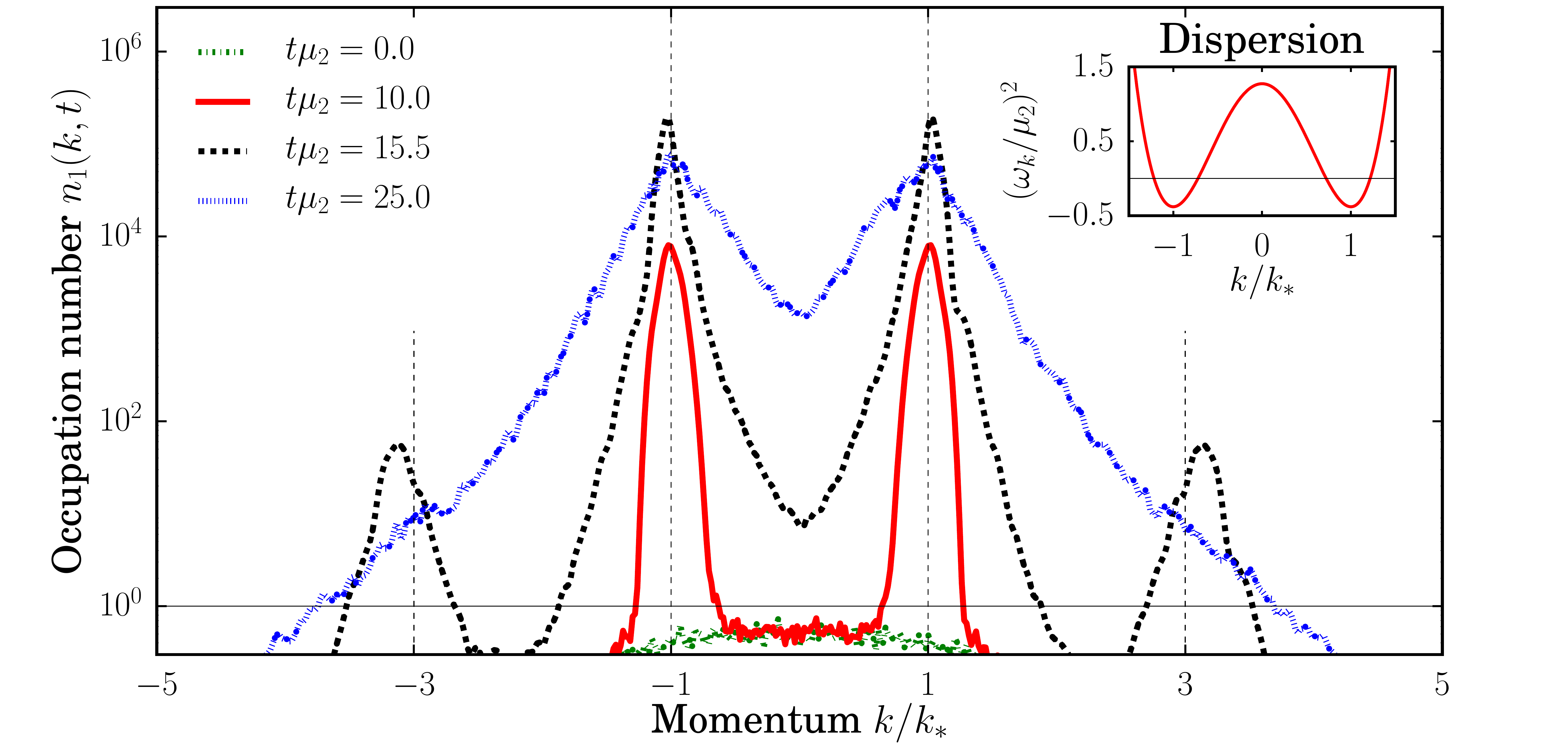}}
	\caption{\label{field1_harmonic_unrealistic}Spectrum of the occupation number $n_1$ at different times. We clearly observe the three regimes of primary growth, secondary instabilities and non-perturbative dynamics (see text). Inset: squared dispersion relation \eqref{disp_rel}. The vertical lines indicate the position of the primary resp. secondary instability at $\pm k_*$ resp. $\pm 3 k_*$.}
\end{figure*}

After having determined the dynamics at early times, which are governed by the dispersion relation \eqref{disp_rel}, we proceed with a numerical simulation. This allows us to access subsequent nonlinear dynamics including secondary growth rates and to validate our analytical calculations for the primary instabilities.

We employ the classical statistical or truncated Wigner approach (TWA) to get more insight into the fluctuations of the atoms in the excited as well as the ground state. To this end, we consider the classical equations of motions
\begin{subequations}
\begin{align}
i \partial_t \varphi_1 &= \left(\!\! -\frac{\nabla^2_z}{2M} + V(z) - \mu_1
+ g_1 |\varphi_1|^2\right) \varphi_1 \notag \\ &\quad+ g_{12} \left(  \varphi_1^{\ast} \varphi^2_2 
+ 2 \varphi_1 |\varphi_2|^2 \right) \, , \label{GP1}\\
i \partial_t \varphi_2 &=\left(\!\! -\frac{\nabla_z^2}{2M} 
+ V(z) -\mu_2 + g_2 |\varphi_2|^2\right) \varphi_2 \notag \\&\quad+ g_{12} \left(  \varphi_2^{\ast} \varphi^2_1 + 2 \varphi_2 |\varphi_1|^2 \right)\, . \label{GP2}
\end{align}
\end{subequations}
We choose a coherent field in the excited state corresponding to the solution $\varphi_{2,s}$ and initialize the lower state with quantum noise. Our main
observable will be the mode occupation number defined as the ensemble-averaged absolute value of the Fourier transformed fields, i.e. $ n_a (k,t) = \braket{\left|\varphi_a(k,t)\right|^2}_W$. The subscript denotes averaging of the classically evolved fields with respect to different initial conditions, which are sampled according to the initial Wigner distribution
\begin{align}
W(\varphi_1, \varphi_2) = \frac{4}{\pi^2} \exp \left(-2 |\varphi_1|^2 - 2 |\varphi_2 - \varphi_{2,s}|^2\right) \; .
\end{align}  
The mode occupation number is connected to the conserved particle number
\begin{align}
N =  \sum_k \left[n_1(k,t) + n_2(k,t) \right] \; .
\end{align}
The classical-statistical approximation is essentially valid for high typical occupation numbers. For a detailed investigation in the case of one-dimensional bosonic gases, see Ref.~\cite{berges2007quantum}.

Figure \ref{field1_harmonic_unrealistic} presents results of a simulation with a harmonic trapping potential (case $a)$). The detailed parameter sets for each simulation can be found in appendix \ref{appendix_parameter}. Shown is the occupation number of $\varphi_1$ as a function of momentum in units of the estimated most unstable mode $k_*$. The corresponding dispersion  is plotted in the inset.

The figure gives results at different times. At initialisation, there are only quantum fluctuations present. However, the system features a dynamical instability which manifests itself in the growth of modes close to $k_*$. At early times the momentum range of the instability is in excellent agreement with our analytical estimate as can been seen from the squared dispersion relation (\ref{disp_rel}) displayed in the inset. At some intermediate time, additional modes around $3 k_*$ become occupied. These momentum modes are initially stable, and they start growing during a secondary stage of amplification after the  primary instability led to sizeable nonlinearities. 

The process of the generation of these secondaries is depicted in figure \ref{secondary_pictorial} and can be understood as follows: The system is prepared in an unstable state, where the whole condensate is transferred to the excited state $\varphi_2$. Consequently, the condensate decays into the energetically favourable state $\varphi_1$. By symmetry this decay occurs as the emission of atom pairs in opposite directions, i.e.\ twin beams with momentum $\pm k_*$. Since the occupation number of the unstable modes grows exponentially, they become highly occupied and the system enters a regime where their self-interaction is no longer negligible. At this point the Bogoliubov approximation breaks down and the stage of secondary amplification sets in.
Processes involving quartic self-interactions between modes then give rise to a particularly fast growth of the mode with momentum $3 k_*$, since the process $\left(k_*, k_*\right) \rightarrow \left(- k_*, 3 k_*\right)$ involves the maximum number of already highly occupied modes allowed by momentum conservation.

\begin{figure}
\centering{
\includegraphics[width= 0.5\textwidth]{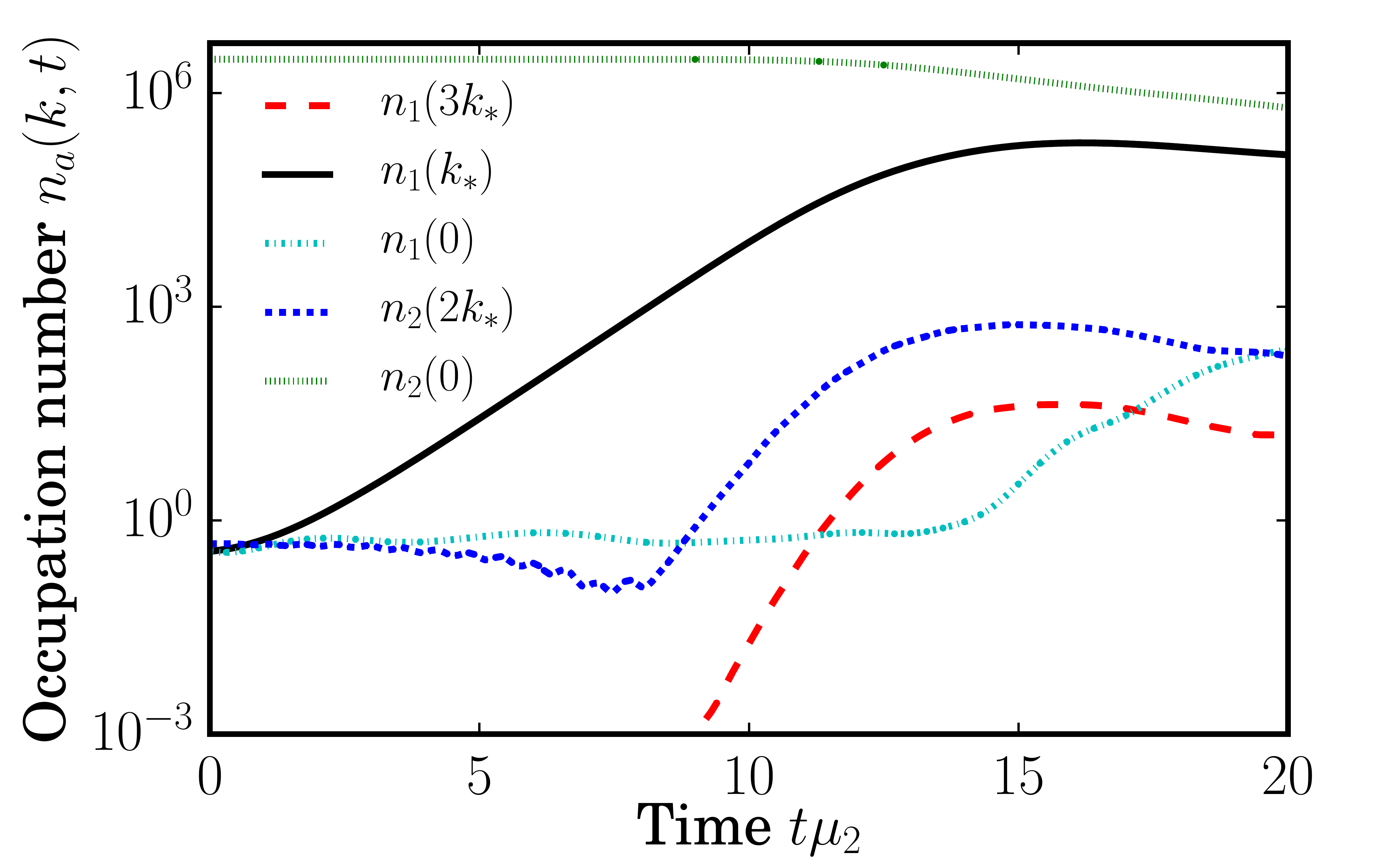}
\caption{\label{growth_unrealistic}Growth of specific mode occupation numbers. Apart from the primary instability for the field mode $\varphi_1(k_*)$, one  finds indications for two different secondaries, $\varphi_2(2k_*)$ and $\varphi_1 (3k_*)$, with larger growth rates as explained in the text. Here straight lines indicate exponential growth due to the logarithmic scale.}}
\end{figure}

Furthermore, the Hamiltonian \eqref{eq:Hamiltonian1} contains another possibility: Two twin beam atoms with momentum $k_*$ can interact with the condensate mode in $\varphi_2$ which is expected to increase the occupation number of the $2k_*$ mode in $\varphi_2$. This is indeed the case as can be seen in figure \ref{growth_unrealistic}, where we have plotted the relevant mode occupation numbers as a function of time.

We emphasize that the occupation numbers are obtained from an equal-time correlation function of two fields. In frequency space such a quantity involves an integral over all possible frequencies and the non-equilibrium processes are not restricted by ``on-shell" energy conservation. Though we are not dealing with a homogeneous system, and momentum conservation can hold at most approximately, corresponding observations are also made for the spatially translation invariant case relevant for cosmology~\cite{berges2003parametric}.  

The simulations reveal that the condensate stays approximately constant for early times, which justifies the Bogoliubov approximation of section \ref{sec:3} to understand the initial instability. As expected, the primary unstable mode grows exponentially and we give a more refined explanation for the secondary growth rates in terms of Feynman diagrams in the next section \ref{sec:5}. For now, we note that the growth rates are higher than those from the primary instability -- the secondaries start later but speed up a lot. As soon as the growth of all modes stops, one enters a nonperturbative regime where the dynamics is no longer dominated by single scattering processes. In figure \ref{field1_harmonic_unrealistic}, we can see that the clear peak structure of the spectrum is washed out in this regime. One also observes from the growth of the zero-momentum mode $n_1(0)$ an indication for the onset of the out-of equilibrium formation of a quasi-condensate in $\varphi_1$ (see figure \ref{growth_unrealistic}).

\section{Diagrammatic Interpretation \label{sec:5}}
In order to make analytical progress in the non-linear regime, we consider the time evolution of the so-called statistical two-point correlation function
\begin{align}
F_{\alpha \beta}(t,z;t',z') &= \frac{1}{2} \left\langle \lbrace \varphi_\alpha(t,z), \varphi_\beta(t',z')\rbrace \right\rangle \nonumber\\
&\quad - \left\langle\varphi_\alpha(t,z) \right\rangle \left\langle\varphi_\beta(t',z') \right\rangle \; ,
\end{align} 
where $\alpha, \beta = 1,2,3,4$ label $\varphi_1, \varphi_1^\dagger, \varphi_2, \varphi_2^\dagger$, respectively. Here $\lbrace A, B \rbrace \equiv AB + BA$ denotes the anti-commutator, which is the appropriate object to consider in the quantum theory because expectation values calculated from the classical-statistical simulations always concern symmetrized field operator products~\cite{berges2007quantum}. Furthermore, the anti-commutator is experimentally accessible since it is related to our main observables $n_a(k,t)$ via
\begin{align}
n_1(k,t) + 2\pi \delta(0) |\bar{\phi}_1^2| = \int dz \; e^{-ikz} F_{12}(t,z;t,0) \; ,\\
n_2(k,t) + 2\pi \delta(0) |\bar{\phi}_2^2| = \int dz \; e^{-ikz} F_{34}(t,z;t,0) \; ,
\end{align} 
where $\bar{\phi}_a$ denotes the condensate part and $n_a$ the respective non-condensate fractions. Introducing also the commutator or so-called spectral fucntion as
\begin{align}
\rho_{\alpha \beta}(t,z;t',z') &= i \left\langle \left[ \varphi_\alpha(t,z), \varphi_\beta(t',z')\right] \right\rangle \; ,
\end{align}
one can derive a set of quantum evolution equations for the nonequilibrium time-evolution of two-point correlation functions~(see e.g.~Ref.~\cite{Berges:2015kfa}). For
given symmetrized part of the self-energy, $\Sigma^F(F,\rho)$, and anti-symmetrized part, $\Sigma^\rho(F,\rho)$, they read   
\begin{widetext}
\begin{align}
\mathcal{D}_{\alpha \gamma}(t,z) F_{\gamma \beta}(t,z;t',z') &= - \int_{t_0}^{t} ds \int dy \; \Sigma_{\alpha \gamma}^\rho(t,z;s,y) F_{\gamma \beta}(s,y;t',z') + \int_{t_0}^{t'} ds \int dy \; \Sigma_{\alpha\gamma}^F(t,z;s,y) \rho_{\gamma \beta}(s,y;t',z') \label{2pi1} \; ,\\
\mathcal{D}_{\alpha \gamma}(t,z) \rho_{\gamma \beta}(t,z;t',z') &= -\int_{t'}^{t} ds \int dy \; \Sigma_{\alpha \gamma}^\rho (t,z;s,y) \rho_{\gamma \beta}(s,y;t',z') \label{2pi2} \; ,
\end{align}
\end{widetext}
where the differential operator $\mathcal{D}_{\alpha \gamma}$ is given in appendix~\ref{app:evol_eqs} along with a more detailed discussion of the equations. 
The self-energies $\Sigma^F$ and $\Sigma^\rho$ contain all interaction contributions such as direct scattering processes or decays. For known self-energies the equations (\ref{2pi1}) and (\ref{2pi2}) are exact and  equivalent to Kadanoff-Baym or Schwinger-Dyson equations~\cite{Berges:2015kfa}. 

The primary instability results in an exponential growth of the statistical two-point function in Fourier space as $F_{\alpha \beta}(t,t',k_*) \sim e^{\gamma_* (t+t')} \quad (\alpha, \beta = 1,2)$. Thus we can estimate nonlinear effects from a resummed loop expansion of the self-energies appearing in the quantum evolution equations \eqref{2pi1} and \eqref{2pi2}. For the non-relativistic Bose gases in one dimension, this is an expansion in the dimensionless diluteness parameter $\zeta = \sqrt{na_s}$ with the scattering length $a_s$ and the average density $n$. In the present case, we have $\zeta \sim {\mathcal O}(\SI{e-5}{})$ depending on the precise parameters of the scenarios discussed in this work and for sufficiently short times, before the typical occupancies become of order $\sim 1/\zeta$, it represents an accurate approach for the description of the dynamics. 

To explain the phenomenon of secondary instabilities, it is sufficient to consider the series up to two loops. In the following, we show how the secondaries can be given an interpretation in terms of Feynman diagrams. The analysis follows along the lines of reference~\cite{berges2003parametric} for the inflaton decay in a cosmological context, which is also reviewed in~\cite{Berges:2015kfa}. We emphasise that the analytical results we present will only provide a  ``leading-log" estimate for the exponential growth rates. A quantitative analysis along these lines including also field amplitudes would require a numerical treatment. 

To determine the secondary growth rates it is sufficient to consider the primary growth of the $1,2$ components of the statistical two-point function in Fourier space as
\begin{align}
F_{\alpha \beta}(t,t',k) \sim e^{\gamma(k)(t+t')} \equiv f_*(t,t',k), \quad (\alpha, \beta = 1,2)
\end{align}
with $\gamma(k) = \sqrt{\gamma_*^2  - \left(\frac{k^2-k_*^2}{2M}\right)^2}$, see \eqref{tree-level prop}. The leading one-loop correction to the evolution equation of $F_{\alpha \beta}$ with $\alpha, \beta=3,4$, i.e.\ the excited field, comes from the self-energy $\Sigma^F$. Thus we neglect all terms of quadratic order in $\rho$, which is justified for $F^2 \gg \rho^2$ for typical modes. 
In momentum space, the corresponding contribution to the RHS of \eqref{2pi1} is given by the memory integral
\begin{align}
&\sim \int_{t_0}^{t'} ds \int dq \; f_*(t,s;q)f_*(t,s;k-q) \rho_{\alpha \beta}(s,t';k)\\
&=\int_{t_0}^{t'} ds \int dq \; e^{\left(\gamma(q)+\gamma(k-q)\right)(t+s)}\rho_{\alpha \beta}(s,t';k) \, .
\end{align}
Due to the exponential growth, the latest times dominate the time integral. Consequently, we approximate $f_*(t,s,p) \approx f_*(t,t',p)$ 
and \begin{align}
\rho_{\alpha \beta}(s,t';k) &\approx \rho_{\alpha \beta}(t',t';k) = \sigma^1_{\alpha, \beta} \quad (\alpha, \beta =3,4) \; ,
\end{align}
where we have used the equal time commutation relations. Furthermore, we simplify the memory integral to include only recent times with $\int_{t_0}^{t'} \mapsto \int_{t'-c/\mu_2}^{t'}$. For an appropriate choice of $c$, the resulting error for the growth rates will be $\sim \log c$. With this approximation we obtain
\begin{align}
\int_{t_0}^{t'} ds \; \rho_{\alpha \beta}(s,t';k) \approx \sigma^1_{\alpha \beta}\int_{t'-c/\mu_2}^{t'} ds \;  = \frac{c}{\mu_2}\sigma^1_{\alpha \beta} \, .
\end{align}
Turning to the momentum integral, we note that it is dominated by $\gamma(k_*) = \gamma_*$ and consider a saddle-point approximation using
\begin{align}
\gamma(k) \approx \gamma_* - \frac{k_*^2}{2 M^2 \gamma_*} (k-k_*)^2 \; ,
\end{align}
where we assume $k_* > 0$ and $\frac{k_*^2}{2M} > \gamma_*$ for simplicity.
At $k=2k_*$ we obtain
\begin{align}
\int dq \; e^{\left(\gamma(q)+\gamma(2k_*-q)\right)(t+t')} \approx \sqrt{\frac{\pi}{\gamma_*(t+t')}} \frac{M \gamma_*}{k_*}e^{2\gamma_* (t+t')} \; ,
\end{align}
which is valid for $\gamma_* (t+t') \gg \left(\frac{M \gamma_*}{k_*^2}\right)^2$. Reinstating all constants, we find the approximation
\begin{align}
F_{\alpha \beta}(t,t';2k_*) \approx  \sigma^1_{\alpha \beta} \frac{g_{12}^2}{2g_2} \frac{c}{\mu_2} \frac{M \gamma_*}{k_*} \sqrt{\frac{\pi}{\gamma_*(t+t')}} e^{2\gamma_* (t+t')} \; ,
\end{align}
which yields the growth rate estimate
\begin{align}
\log F_{34}(t,t;2k_*) \approx 4 \gamma_* t \; .
\end{align}
In a similar fashion, we can estimate the leading contribution to the $\alpha, \beta =1,2$ components to be
\begin{align}
\log F_{12}(t,t;3k_*) \approx 6\gamma_* t \; .
\end{align}
\begin{figure}
\centering{\includegraphics[scale=1.1]{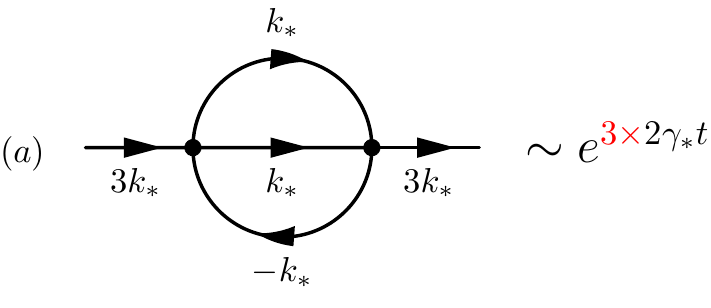}
    \includegraphics[scale=1.1]{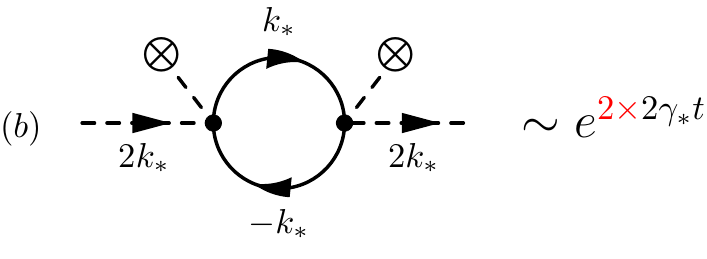}
    \caption{\label{feynman} Feynman diagrams associated to the secondary growth rates. The solid (dashed) lines correspond to the propagators of $\varphi_1$  ($\varphi_2$). The circled crosses denote interactions with the quasi-condensate in $\varphi_2$. The estimated growth rates directly reflect the number of propagator lines of the most unstable $\pm k_*$ present in the loop diagrams. An approximate computation of $(b)$ is given in the main text.}}
\end{figure}
\begin{figure}
\centering{\centering{\includegraphics[scale=0.2]{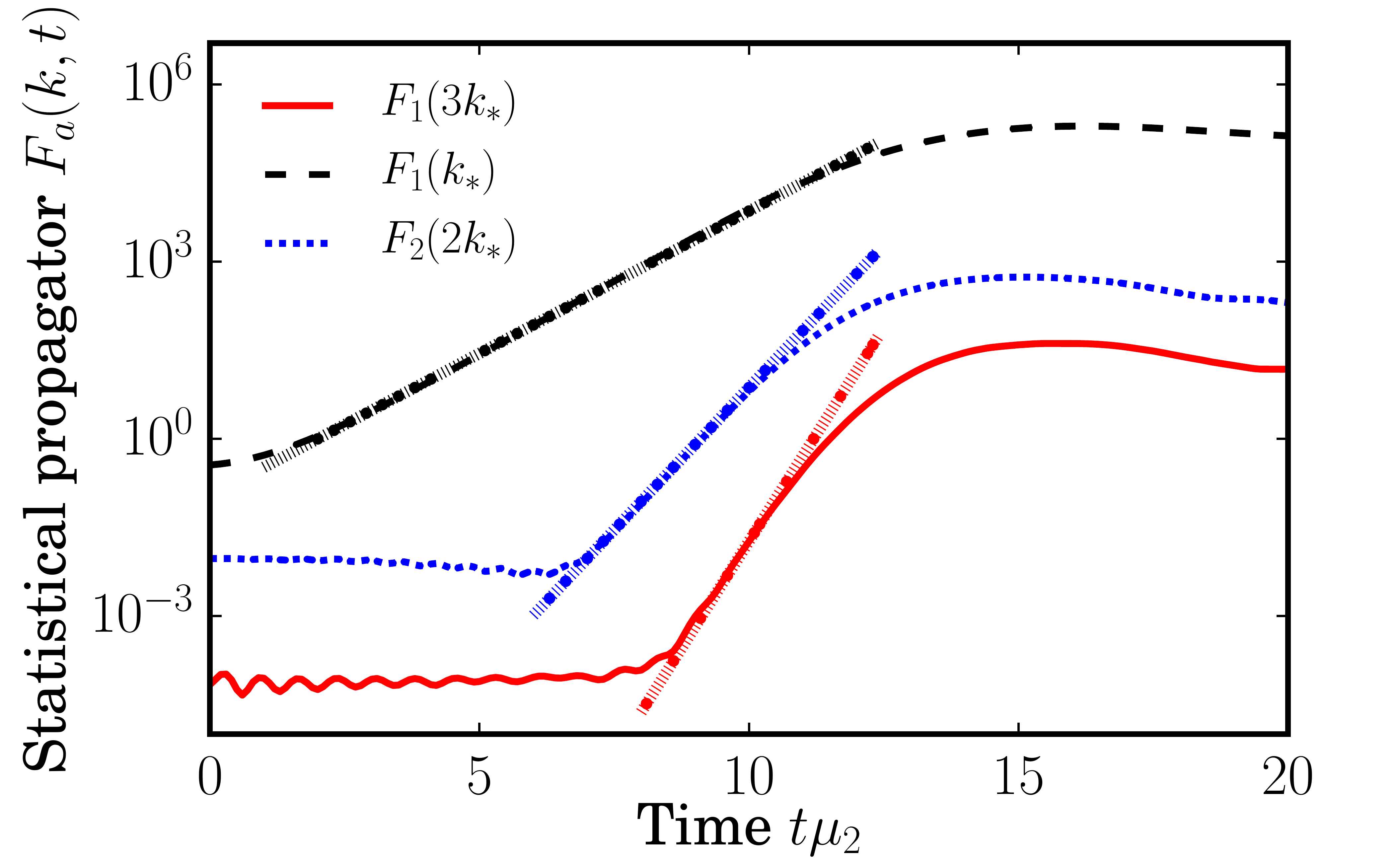}}
\caption[Simulation 1: The statistical propagator]{\label{stat_prop}Time evolution of primary and secondary unstable Fourier modes of the statistical propagator $F_a(t,k) = \int dz \; e^{-ikz} \frac{1}{2}  \left\langle \left\lbrace \phi_a(t,z) , \phi_a^\dagger(t,0) \right\rbrace \right\rangle_\text{c}$. The straight lines indicate exponential growth. The black, long-dashed curve shows the primary instability with the reduced primary growth rate $2\bar{\gamma}_1 \approx 0.9 \times 2\gamma_*$. The blue, dotted resp. red, solid curves are the secondary unstable modes. We observe the secondary growth rates to agree initially with the estimated rates $2\bar{\gamma}_j \approx j \times 2\bar{\gamma}_1$ for $j = 2$ resp. $j=3$.}}
\end{figure}
The general interpretation is depicted in figure \ref{feynman}, where we show the two- and one-loop diagrams that can be associated with the generation of secondaries. In this simple picture, the number of internal propagator lines encodes which modes become unstable as well as how fast they grow. 

Since for systems out of equilibrium different loop diagrams can become important at different times, we consider a dynamical power counting and give parametric estimates for the time-scales of the different diagrams next. A general loop diagram containg $n$ vertices, $m$ propagator lines and $k$ condensate fields will scale parametrically as $\sim \zeta^{n-k} F^m$. In accordance with the above estimates, we consider diagrams that contain the most unstable mode as internal propagator lines, since these contributions are expected to dominate the dynamics.

At first loop-order, we have the diagram \parbox{1cm}{\includegraphics[scale=1]{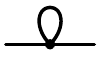}} for the lower field $\varphi_1$. This will play an important role in the dynamics as soon as it becomes parametrically of order one. We can estimate the time-scale by  $1 \simeq \zeta e^{2 \gamma_* t}$, resulting in
\begin{equation}
t_3 \simeq \frac{1}{2 \gamma_*} \log \left( \zeta^{-1}\right) \, .
\end{equation}
This is a conservative lower bound since we have assumed the most unstable exponential growth for all modes which is somewhat overestimating the effect.
Around $t_3$, we find sizeable deviations from the behaviour of the most unstable  and of the condensate mode. As a consequence, there are corrections of order one which come from diagrams with an arbitrary number of loops. Thus the dynamics is no longer characterized in terms of the small expansion parameter $\zeta$. This signals a nonperturbative regime as indicated above (see figure \ref{field1_harmonic_unrealistic}).

At second order in $\zeta$, we have the setting-sun diagram  \parbox{1.5cm}{\includegraphics[scale=1]{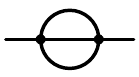}}  for the lower field $\varphi_1$ shown in figure \ref{feynman}$a)$. Proceeding as for the previous diagram, we use $1 \simeq \zeta^2 e^{6 \gamma_* t}$ to find the parametric estimate for the time-scale
\begin{equation}
t_2 \simeq \frac{2}{3} t_3 \, .
\end{equation} 
Remarkably $t_2 < t_3$, i.e.\ the two-loop correction becomes important \emph{earlier} than the previous one-loop diagram. This two-loop process can be associated with the generation of secondaries. Employing momentum conservation, three propagator lines with $k_*$ combine to $3 k_*$ and give rise to a secondary growth rate of $6 \gamma_*$.

Similarly, we can estimate the one-loop correction  \parbox{1.5cm}{\includegraphics[scale=1]{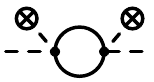}} for the excited field $\varphi_2$ to become important at the time
\begin{equation}
t_1 \simeq \frac{1}{2} t_2
\end{equation}
which follows from $1 \simeq \zeta^2 \left(\zeta^{-1/2}\right)^2 e^{4 \gamma_* t}$. It is important to note that the diluteness parameter enters twice here, once for each vertex and as $\sim \zeta^{-1/2}$ for the non-vanishing field, such that the diagram is effectively of order $\zeta$. This process can be associated with the secondary growth at $2 k_*$ in $\varphi_2$ with rate $4 \gamma_*$ calculated above (see figure \ref{feynman}b).

Our findings for the time-scales $t_1 < t_2 < t_3$ agree well with the time evolution of the statistical two-point function plotted in figure \ref{stat_prop}. The secondary at $t_1$ in $\varphi_2$ starts before the one at $t_2$ in $\varphi_1$ and finally the system becomes non-perturbative at $t_3$. 
The observed growth rates are lower than predicted because nearby modes with different rates $\gamma(k) < \gamma_*$ contribute and effectively damp already the primary instability. In order to test our estimates for the secondary rates, we correct for this damping by choosing the largest pointwise derivative for the primary rate and find $\bar{\gamma}_1 \approx 0.94 \times \gamma_*$, which is slightly smaller than $\gamma_*$ as expected due to the harmonic trapping potential. Similarly, the secondary rates are estimated from the numerics as $\bar{\gamma}_2 \approx 1.86 \times \gamma_* \approx 1.98 \times \bar{\gamma}_1$ and $\bar{\gamma}_3 \approx 2.80 \times \gamma_* \approx 2.98 \times \bar{\gamma}_1$. Thus we find  a quantitative agreement of the estimated secondary rates at the onset of the exponential growth after correcting for the systematic error of the primary instability. At later times, the interaction with other modes becomes non-neglible and consequently the approximated loop calculation does not describe the dynamics anymore. Instead the growth eventually stops. As a result, one can only observe the secondaries for one or two orders of magnitude in growth for the statistical two-point correlator.

The above reasoning is very general and also explains why secondaries are a generic feature of seemingly unrelated systems, such as ultracold atoms and cosmological inflation. The important ingredients are a primary instability, such as the tachyonic one considered in this work, and a four-vertex of the model Hamiltonian. In this sense the present system provides a minimal model to study non-linear phenomena related to secondaries. In particular, the mechanism is insensitive to the system being non-relativistic or relativistic as in the case of inflation.

\begin{figure}
\centering{\includegraphics[width= 0.5 \textwidth]{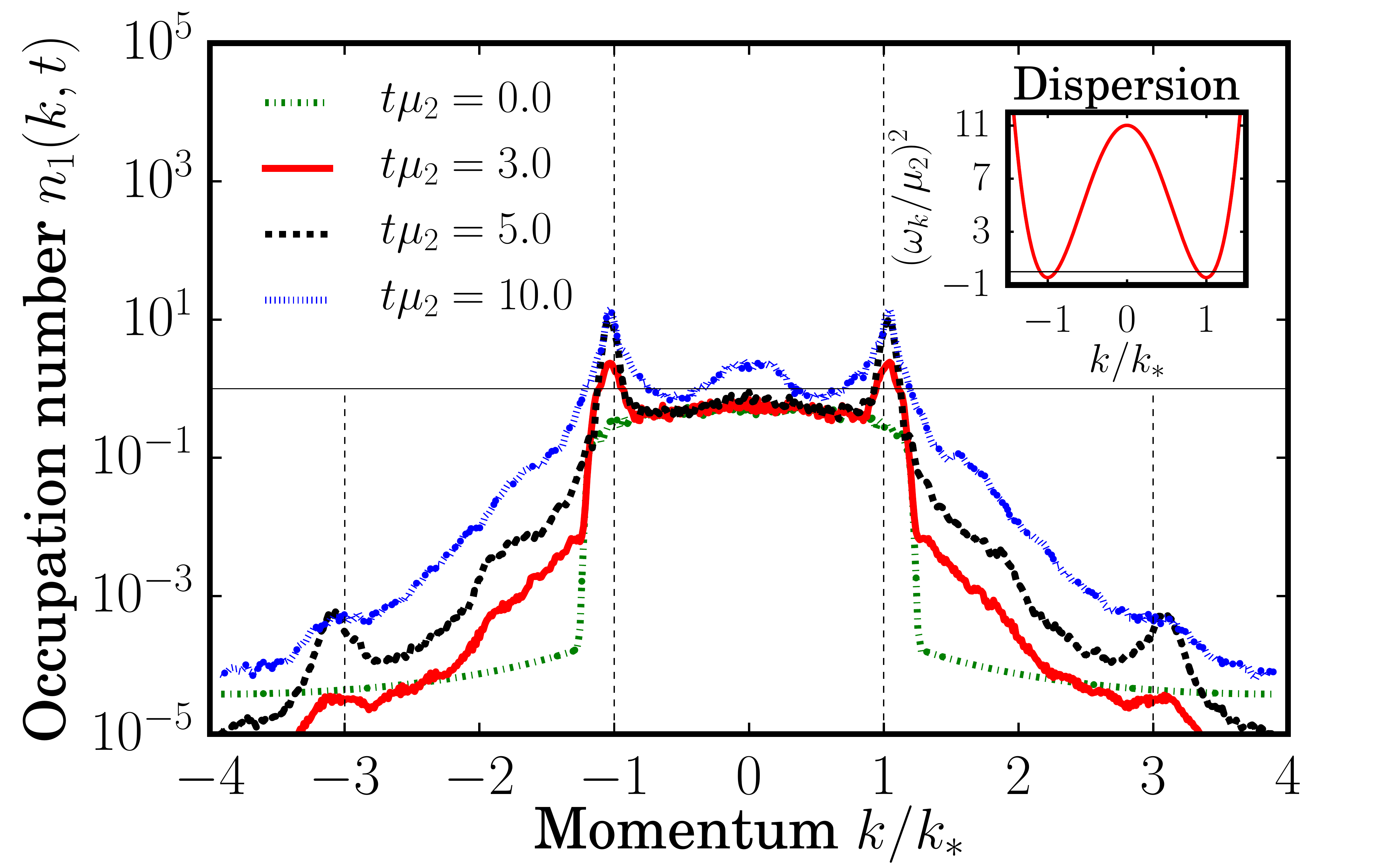}}
\caption{\label{realistic_sim} Occupation number distribution for $\varphi_1$ using parameters characteristic of the twin beam experiment \cite{bucker2011twin}. The secondaries are present but  beyond experimental reach as indicated by the black horizontal line. The inset shows the squared dispersion relation.}
\end{figure}

\section{Secondary Instabilities with Seed \label{sec:6}}
 In principle, the process of secondaries repeats itself and leads to the occupation of even higher momentum modes. It is this amplification mechanism that is proposed to enhance inflationary particle production over a wide range of momenta. It would be fascinating to test this non-linear aspect of inflaton dynamics in a real cold-atom experiment. Unfortunately, our simulations suggest that it is difficult to detect this effect in a realistic setup without further optimization. 
 
 To be specific, we have simulated the twin beam experiment of \cite{bucker2011twin} for characteristic parameters and plotted the result in figure \ref{realistic_sim}. 
The spectrum of $\varphi_1$ exhibits twin beams which have also been observed in experiment. The green dashed-dotted line shows the range of initial quantum fluctuations which we have cut off next to the primary instability. We have explicitely checked that the twin beams are insensitive to the choice of the cutoff as long as the fluctuations are present within the range of unstable modes. Our choice reveals the possible excitation of secondaries at the expected $\pm 3 k_*$. They have not been observed in the experiment, probably because of the small occupation number compared to fluctuations which is due to particle number limitations ($N \approx 800$). The naive ansatz of increasing the number of available atoms does not solve the problem because it increases the width $\delta k$ of the primary instability. At the same time $k_*$ shifts to lower momenta and the instability eventually vanishes (see equations \eqref{fineness} and \eqref{primary}). This results in a non-trivial optimization problem limited by accessible setup parameters.

\begin{figure*}
\centering{
\includegraphics[width= 0.5 \textwidth]{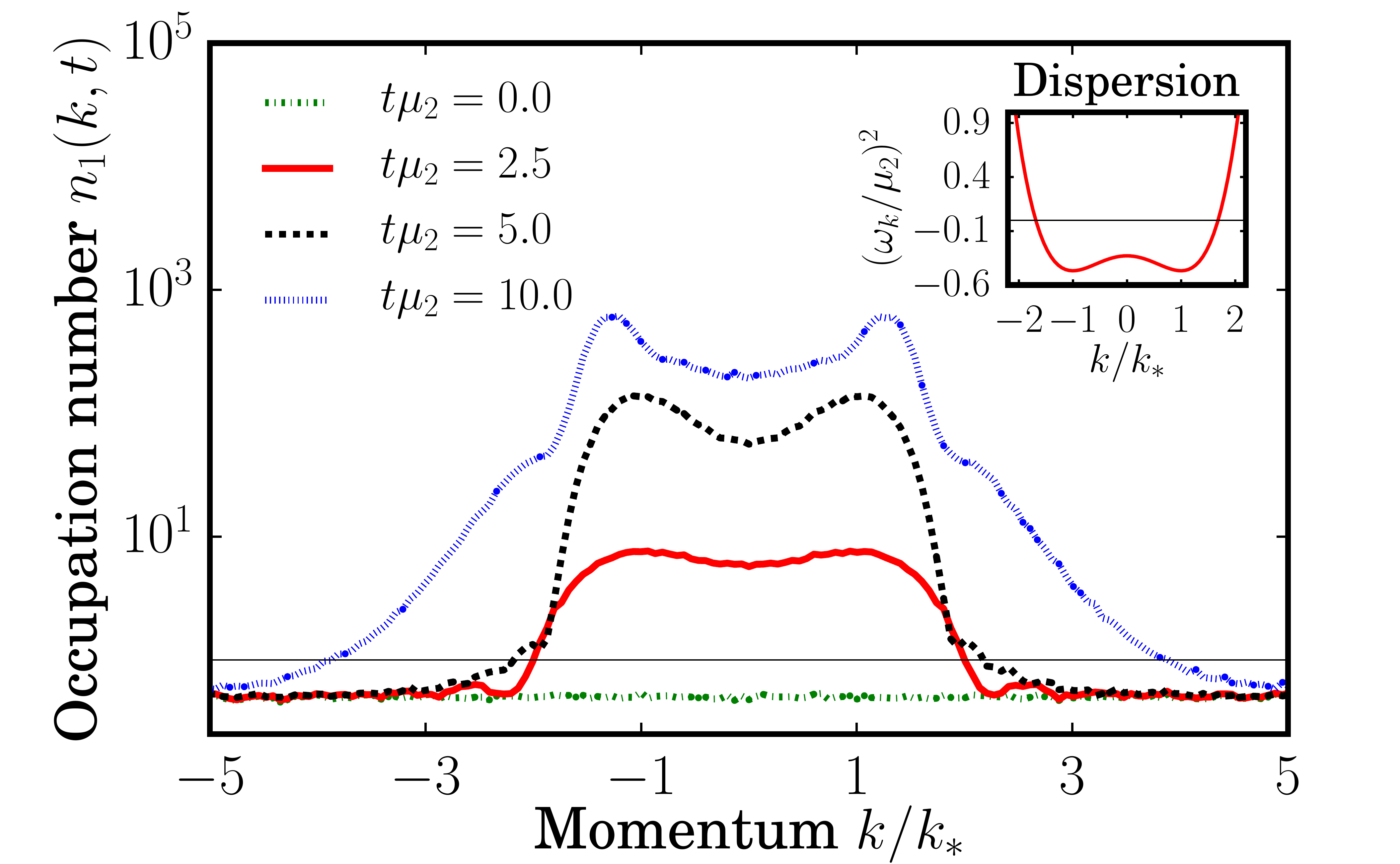}\includegraphics[width= 0.5 \textwidth]{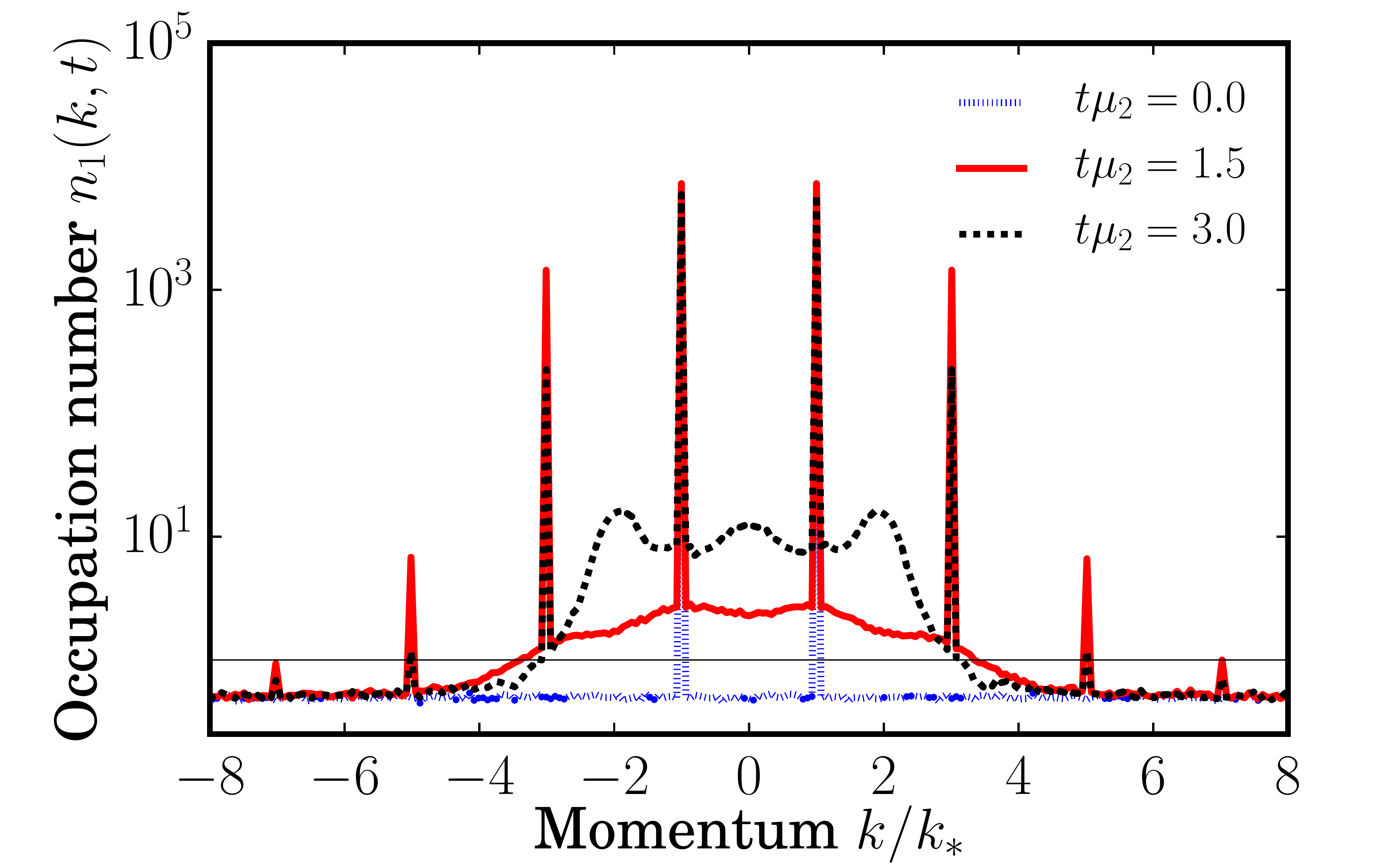}}
\caption{\label{seed-noseed}Occupation number distribution for $n_1$. Left: Without seed, we only observe a very broad initial instability (inset: squared dispersion relation). Right: With seed, nonlinear interactions immediately lead to secondaries characterized by clear peaks at odd multiples of the seeded mode $k_*$. Note that we display earlier times and larger momenta due to the faster  dynamics redistributing the occupation numbers.}
\end{figure*}

The above results suggest that the most important ingredient for the generation of secondaries is a high occupation of the primary instability. Thus, we propose to seed the initial instability significantly. Instead of preparing the whole condensate in the excited state $\varphi_2$, we keep a fraction of all atoms in the lower state $\varphi_1$ and imprint a modulation to transfer them into the most unstable modes $\pm k_*$, i.e.\ we choose a coherent initial field value of
\begin{equation}
\varphi_1 (z, t=0) = \beta \varphi_{2,s}^{(b)}(z) \left(e^{ik_*z} + e^{-ik_*z}\right)
\end{equation}
where the fraction $0 < \beta < 1$ controls the seeding strength. Tuning the seed, we expect to effectively skip the regime of primary growth and immediately observe secondaries. Furthermore we have often employed momentum conservation in the previous section. Strictly speaking this is not given in the presence of a harmonic trapping potential (case $a)$). Recently, almost perfect box potentials have been achieved experimentally \cite{gaunt2013bose}. Thus we may eliminate the systematical error of not having momentum conservation by choosing box-type potentials (case $b)$) in the following. Figure \ref{seed-noseed} shows the direct comparison of the two scenarios with and without seed.

The situation without seed is similar to the twin beam case. Comparing the early occupation (in red) to the squared dispersion relation (inset) again demonstrates the validity of the Bogoliubov approximation. However, this time the primary instability is very wide which is due to a different parameter set and mainly results from a larger particle number. At later times, the twin beams wash out and we find again the spectrum that we have previously associated with non-perturbative dynamics. 

\begin{figure}
\centering{\includegraphics[width= 0.5 \textwidth]{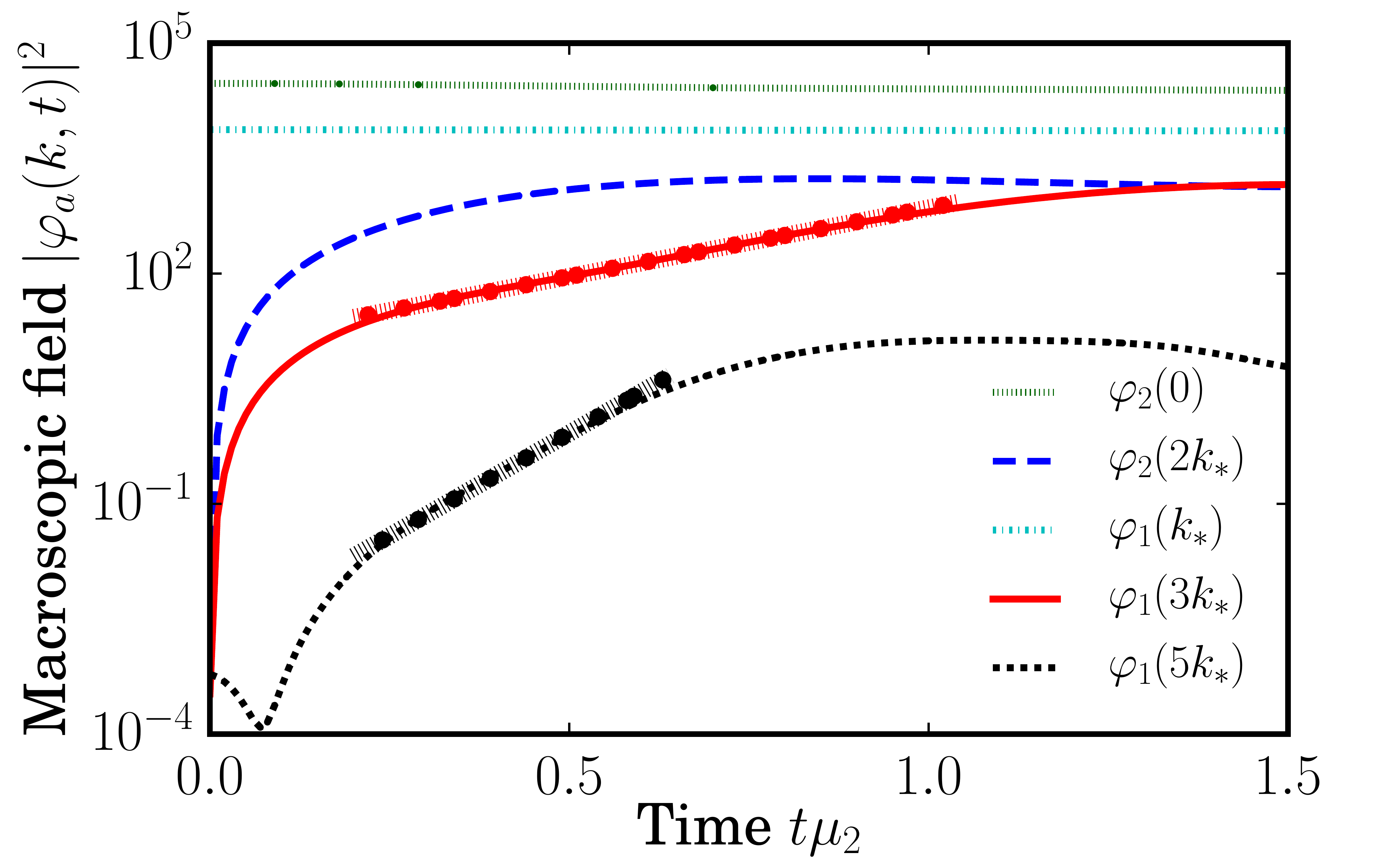}}
\caption{\label{hom_growth} Growth of the squared absolute value of the macroscopic field $|\varphi_a(k,t)|^2$ for the seeded simulation. The condensate $\varphi_2(0,t)$, as well as the seeded mode $\varphi_1(k_*,t)$ show no significant dynamics, while the secondaries in the modes $\varphi_1(3k_*,t), \varphi_2(2k_*, t), \varphi_1(5k_*,t)$ grow fast. The straight lines indicate exponential growth with a rate of $3\gamma_*$ (red) and $9 \gamma_*$ (black)}
\end{figure}

This is to be compared to the situation with seed: The primary instability is highly occupied early on and consequently the Bogoliubov dispersion does not dominate the dynamics. Instead we observe clear peaks at odd multiples of the seeded mode. They are created from a cascade of scattering events as explained in section \ref{sec:4}. Moreover the underlying dynamics is much faster than the primary growth such that the latter is suppressed in the experiment. One could argue that we are merely observing a tree level scattering process here. That we are dealing with an amplified nonlinear process can be verified via the growth rate of the secondary instabilities, which are shown in figure \ref{hom_growth}.
Due to the presence of a large initial one-point function in the lower field $\varphi_1$, the seed shifts the dominant dynamics from the two-point $F$ to the macroscopic field. In fact, we have plotted the absolute value of the latter, where we clearly observe a higher growth rate similar to the unseeded case of the previous section. Moreover, the rate is on the order of $6 \gamma_*$ which we have estimated in the previous section. We postpone the discussion of the observed exponential growth of the $5k_*$ mode until the end of the next section. Furthermore, we observe that the $2k_*$ and $3k_*$ modes show a very peculiar early-time behaviour that was not present in the unseeded case. A doubly logarithmic plot (not shown) reveals that time-dependence follows a power law $\propto t^2$. Since the nonlinear corrections become important only at sufficiently late times $\gamma_*t \gg 1$, a naive explanation for this effect is the following: If the fields admit a Taylor expansion with vanishing constant term, i.e.
\begin{align}
\varphi(t) \approx C + C' \left( \gamma_*t \right) + \mathcal{O}\left((\gamma_* t)^2 \right) \; ,
\end{align}
with $C = \varphi(0) = 0$ and $C' = \frac{\varphi'(0)}{\gamma*}$, then $\left| \varphi(t) \right|^2 \propto t^2$ for $\gamma_* t \ll 1$. We will verify this reasoning in a classical toy model in the next section.

\section{Simplified Four-Mode Model\label{sec:7}}
In the previous section, we have seen that the seeding procedure simplifies the dynamics in the sense that the evolution gets dominated by one-point functions. This motivates us to study the coupled equations \eqref{GP1} and \eqref{GP2} in more detail. We have also seen that the dynamics involve mainly a few modes, namely odd multiples of $k_*$ of the lower field $\varphi_1$ and even multiples of $k_*$ of the excited field $\varphi_2$. Consequently, we transform the coupled equations for box-type potentials to Fourier space and restrict them to $\varphi_{mk_*}^{(1)}$ with $m$ odd and $\varphi_{nk_*}^{(2)}$ with $n$ even. We further restrict ourselves to the first possible secondaries at $2k_*$ and $3 k_*$, i.e.\ $m=\pm 1, \pm 3$ and $n=0, \pm 2$. Employing the symmetry $k \rightarrow -k$, we are left with a system of four coupled non-linear differential equations for the complex functions $\varphi_1 \equiv \varphi^{(1)}_{k_*}(t), \varphi_3 \equiv \varphi^{(1)}_{3k_*}(t), \varphi_0 \equiv \varphi^{(2)}_{0}(t), \varphi_2 \equiv \varphi^{(2)}_{2k_*}(t)$ of the form
\begin{align}
i \partial_t \varphi_j = \mathcal{H}_j\left[\varphi_0, \varphi_1, \varphi_2, \varphi_3\right] \qquad j =0,1,2,3 \; .
\end{align}
The precise form of $\mathcal{H}_j$ is given in appendix~\ref{appendix_parameter}.  
The model has four independent parameters: Three interaction constants $c_1, c_2, c_{12}$ and the dimensionless combination $\frac{k_*^2}{M \mu_2}$. Most importantly, it contains nonlinear terms which introduce momentum-conserving scattering processes between the different modes in accordance with the original Hamiltonian \eqref{eq:Hamiltonian1}. We do not expect to find any numerical agreement between this highly simplified model and the full quantum dynamics. However, we suspect that in the presence of seeds the feature of secondaries can be qualitatively understood with this rather simple differential equation.

The initial value problem that we have considered before corresponds to a high initial occupation of $\varphi_0(t_0) = \sqrt{\frac{\mu_2}{c_2}}$, which is determined by the stationary solution of the equations of motion with all other modes set to zero. All other initial values we set to very small values $\varphi_1(t_0), \varphi_2(t_0),\varphi_3(t_0) \ll \varphi_0(t_0)$. Then a linearised analysis for the most unstable mode yields
\begin{align}
i \partial_t \varphi_1 \simeq \gamma_* \varphi_1^\dagger \; ,
\end{align}
which for the corresponding mode occupation number is solved by $\left| \varphi_1\right|^2 = A e^{2\gamma_*t} + B e^{-2\gamma_* t}$. Here we recover again the Bogoliubov result and notice that a non-vanishing initial occupation is necessary to trigger the primary instability. In the quantum system, these initial fluctuations are always present as the ``quantum-half". To keep the discussion general, we consider an arbitrary initial occupation $\left| \varphi_1(t_0)\right|^2 = A + B = \beta^2 \left| \varphi_0(t_0)\right|^2$. In accordance with our previous simulations, we choose the lower and upper fields to be oscillating in phase, i.e.\ $A - B = 0$. Consequently, the primary instability is given by
\begin{align}
\left| \varphi_1(t)\right|^2 = \beta^2 \left| \varphi_0(t_0)\right|^2 \cosh \left(2 \gamma_* t\right)
\end{align}
with a small (seeding) parameter $0 < \beta < 1$, which justifies the linearization at early times.

If the system shows a clear separation of scales, then the zero mode will stay relatively highly occupied, while the most unstable mode grows exponentially. Thus, we consider only terms with the highest powers of the most unstable mode in the remaining two equations of motions, i.e.\ we approximate
\begin{subequations}\label{sec_eq}
\begin{align}
i \partial_t \varphi_2 &\simeq 3 \beta^2 c_{12} \left| \varphi_{0}(t_0)\right|^3 \cosh \left(2 \gamma_* t\right) \; ,\\ 
i \partial_t \varphi_3 &\simeq \beta^{3} c_1 \left| \varphi_0(t_0)\right|^{3}\cosh^{3/2} \left(2 \gamma_* t\right) \; . 
\end{align}
\end{subequations}
Of course, these equations can only be valid for a certain intermediate regime, where other contributions can be neglected. As we will see below, they lead to exponential growth, such that terms involving the fields themselves will gain importance over time and invalidate the approximation at late times. On the other hand, the initial values of the fields have to be sufficiently small compared to $\varphi_{k_*}^{(1)}(t_0)$ in order for the approximation to be valid early on. The precise regime of validity is rather difficult to estimate. Nevertheless, we will use this approximation to determine the exponential growth rates and justify the approximation afterwards numerically.

In fact, we can directly integrate equations \eqref{sec_eq} as
\begin{align}
\varphi_2(t) &\sim \int dt \; \cosh(2\gamma_*t) \sim  \sinh \left(2 \gamma_* t\right) \; , \label{int1}\\
\varphi_3(t) &\sim \int dt \; \cosh^{3/2}(2\gamma_*t) \\
&\sim \sinh \left(2\gamma_* t\right) \sqrt{\cosh\left(3\gamma_* t\right)} - i F(i\gamma_* t | 2)\;  , \label{int2}
\end{align}
where $F(x | m)$ denotes the elliptic integral of the first kind with parameter $m$. Expanding the solutions at late times $\gamma_* t \gg 1$, we recover the results of the loop calculation, 
\begin{align}
\left|\varphi_2(t)\right|^2 \sim  e^{2 \times 2\gamma_* t} \; , && \left|\varphi_3(t)\right|^2 \sim  e^{{3 \times} 2\gamma_* t} \; .
\end{align}

\begin{figure}
\centering{\includegraphics[scale=0.2]{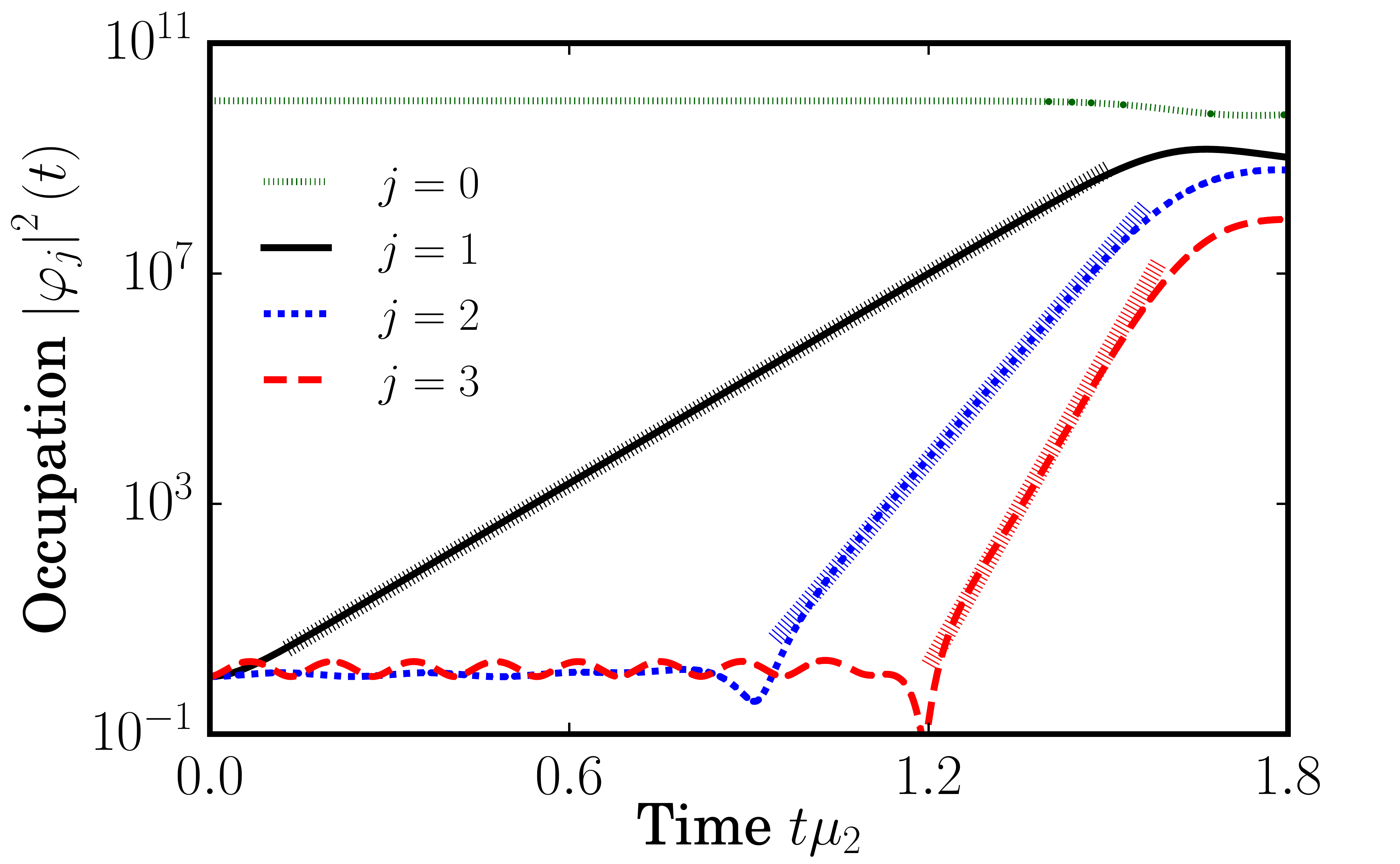}}
\caption[Secondaries within the four-mode model]{\label{4mode_sim}Squared absolute value of the four modes $\varphi_j$ as a measure for the mode occupation number. The straight lines indicate exponential growth with rates $2 \gamma_*$ (solid black), $4 \gamma_*$ (dotted blue) and $6 \gamma_*$ (dashed red).  The precise parameters for the simulation are given in appendix \ref{appendix_parameter}.}
\end{figure}

To substantiate our analytical predictions, we show the results of a numerical simulation of the coupled evolution equations in figure \ref{4mode_sim}. We observe very similar dynamics compared to the statistical propagator of the first TWA simulation (cf. figure \ref{stat_prop}). The primary $\varphi^{(1)}_{k_*}$ exhibits to very good accuracy the expected exponential growth rate of $2\gamma_*$ and even the growth rates of the secondaries $\varphi^{(2)}_{2k_*}$ and $\varphi^{(1)}_{3k_*}$ agree well with the estimates $4 \gamma_*$ and $6 \gamma_*$, respectively. We conclude that the simplified four-mode model encompasses the basic ingredients for the generation of secondaries, i.e.\ a primary instability, the necessary non-linear scattering terms and one highly occupied condensate mode $\left(\varphi^{(2)}_{0}\right)$ serving as a particle bath.

\begin{figure}
	\centering{\includegraphics[scale=0.2]{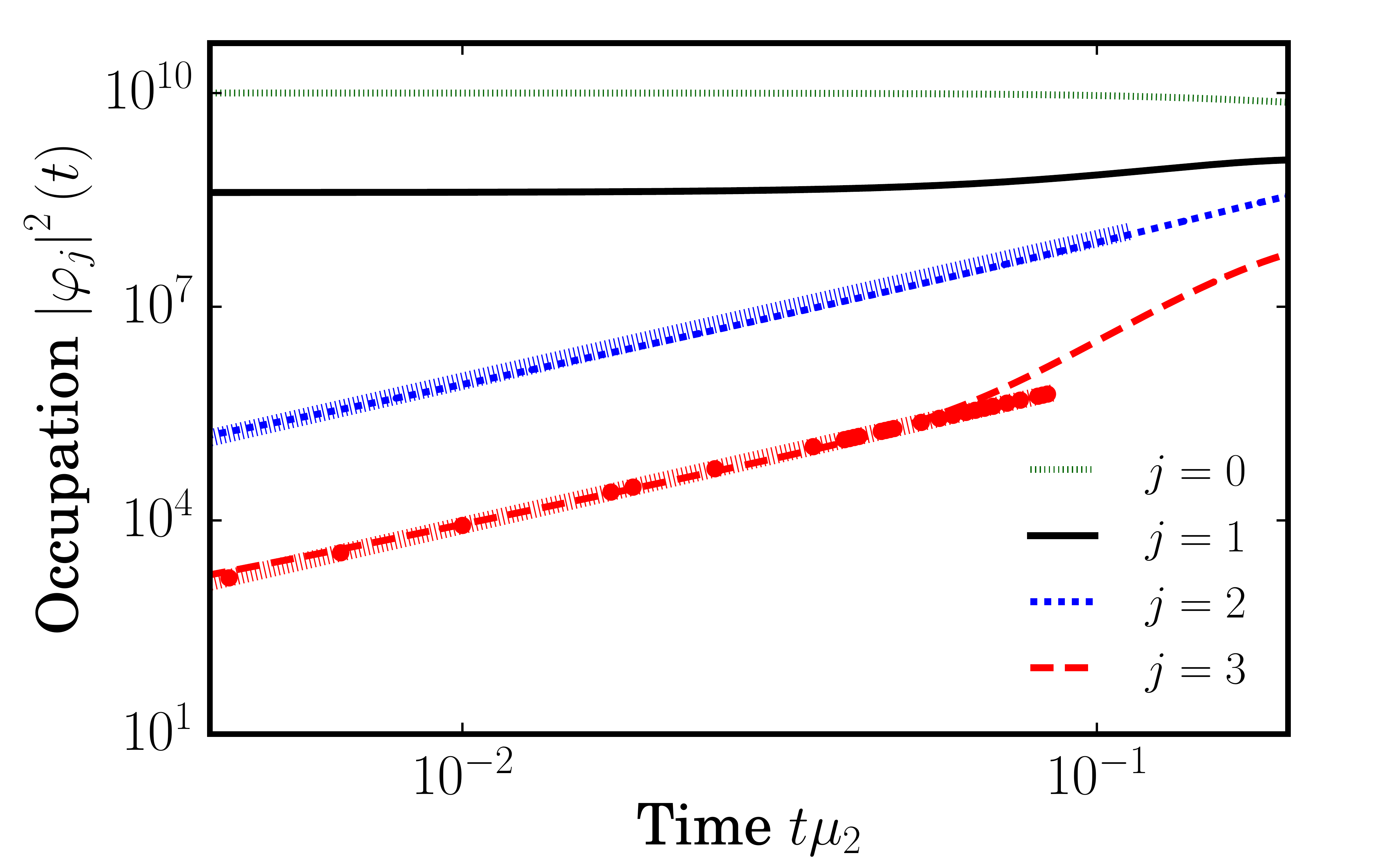}}
	\caption[Early time behaviour within the seeded four-model]{\label{4mode_sim_early}Squared absolute value of the four modes for the same parameters used in the simulation for figure \ref{4mode_sim} except for the primary $\varphi_{k_*}^{(1)}$, whose initial value is set by $\beta = \frac{1}{5}$. The straight lines indicate a power-law $\propto t^2$ on the double logarithmic scale.}
\end{figure}

We now turn to the unexpected short-time dynamics that we have observed in the seeded case. Expanding the solutions \eqref{int1} and \eqref{int2} at early times $\gamma_* t \ll 1$ yields
\begin{align}
\left|\varphi_2(t)\right|^2 \sim \left(4\gamma_* t\right)^2 \; , && \left|\varphi_3(t)\right|^2 \sim \left( 6\gamma_* t\right)^2 \; ,
\end{align}
which explains the observation of a power law $\propto t^2$ seen in figure \ref{hom_growth}. To verify the analytics, we have plotted the results of another simulation in figure \ref{4mode_sim_early}, where we have increased the initial occupation of the primary unstable mode in order to compare to the situation with seed. We cleary observe that the secondaries show initially a growth $\propto t^2$. Additionally, one can still fit an exponential growth to the $3k_*$ mode (not shown in the plot), but not to the $2k_*$ mode, which agrees with the observation of the seeded TWA simulation. The strong seed assures an initially small occupation of the secondary modes compared to the primary, which allows us to make the approximation of equation \eqref{sec_eq} at early times. In conclusion, the naive interpretation of the previous section, given in terms of a Taylor expansion with vanishing time-independent constant, indeed explains the behaviour of the secondaries at early times.

For values of $\beta$ close to one, i.e.\ the strongly seeded situation, the linearised analysis, which we employed to determine the primary instability, may be questioned. Since we cannot solve the system analytically, we can only give some heuristic comments on the effect of seeding: A high initial occupation of the most unstable mode breaks the translational invariance of the linearised equations of motion and leaves a discrete symmetry. One can then apply Bloch's theorem: We consider e.g.\ the secondary growth $\varphi^{(1)}_{3k_*} (t) \propto e^{3\gamma_* t}$ observed in figure \ref{hom_growth}. In the spirit of the simplified few-mode model, this field will couple e.g.\ to the Fourier mode $k = 9 = 3 + 3 + 3$ giving rise to the rate $\varphi^{(1)}_{9k_*} (t) \propto e^{9\gamma_* t}$, which follows from the nonlinear term $\left(\varphi^{(1)}_{(9-3-3)k_*} \right)^* \varphi^{(1)}_{3k_*}\varphi^{(1)}_{3k_*}$ by momentum conservation. However, in the seeded case, this growth rate will already manifest itself in a lower mode because we may identify modes in different Brillouin zones as $k \equiv k \mod 2k_*$. Of course, this influence is suppressed, but this suppression might be compensated by the high occupation numbers. In particular, the nonlinear term $\left(\varphi^{(1)}_{(5-3-3)k_*} \right)^* \varphi^{(1)}_{3k_*}\varphi^{(1)}_{3k_*}$ can be turned into $\left(\varphi^{(1)}_{-3k_*} \right)^* \varphi^{(1)}_{3k_*}\varphi^{(1)}_{3k_*}$ by replacing $-k_* \rightarrow -3k_*$ only once. Thus, also the mode $5k_*$ can show the high growth rate of $9\gamma_*$, which may explain the observation in figure $\ref{hom_growth}$ and concludes our analysis of secondaries.

\section{Conclusion and Outlook\label{sec:8}}

In the present study we have demonstrated that the generic feature of the non-linear secondary growth of unstable modes, which was first predicted in the context of inflationary particle production, prevails in various cold-atom setups. We have analyzed possible choices of realistic parameter sets and indicated why recent twin beam experiments have not detected this effect. The main obstacle concerns particle number limitations. In order to make the secondary growth of unstable modes experimentally accessible, we have proposed a seeding procedure which effectively amplifies the underlying nonlinear corrections. 

We have furthermore demonstrated that the growth rates of secondary amplifications allow an interpretation in terms of loop diagrams, which reflect the structure of interactions of the underlying Hamiltonian. This is a direct macroscopic visualization of (quantum) loop processes. In turn, one may deduce essential aspects of the interaction structure of an unknown Hamiltonian from  measured instabilites. This provides a striking example where macroscopic observables determine fundamental microscopic properties of a many-body system.

In this work, we have focused on the early-time quantum dynamics including primary and secondary growth. Furthermore, we have pointed out a highly simplified model that essentially captures the same dynamical features. Our simulations and analytical estimates suggest that the full quantum model subsequently enters a non-perturbative regime characterized by the importance of diagrammatic contributions with an arbitrary number of loops. This regime cannot be simulated by a simplified few-mode model due to the strongly correlated nature of many interacting degrees of freedom. The $\mathcal{O}(N)$-symmetric inflaton model - showing the corresponding behaviour at early times - exhibits universal scaling dynamics associated to an attractor far from equilibrium during a similar non-perturbative regime \cite{Berges:2008wm,orioli2015universal}. It will be interesting to study the possibility of observing a so-called nonthermal fixed point in a realistic cold-atom experiment. However, one spatial dimension is special due to phase space restrictions and near-integrability. The model considered in this work encorporates two-body scattering exchanging particle types between the two fields. The presence of this interaction can be thought of as arising from breaking the symmetry to a smaller subgroup $\mathcal{O}(4) \rightarrow \mathcal{O}'(2)$, where $\mathcal{O}'(2)$ denotes a combined rotation of the two fields $\phi_{1/2}$. This might facilitate the presence of a universal attractor far from equilibrium. In that context, we would also like to study the influence of the dimensionality of space-time for the universality class in more detail. We reserve a detailed analysis of the long-time evolution of the discussed two-field system for future studies.

\begin{acknowledgments}
We thank T.~Gasenzer, M.~Oberthaler, A.~Pi\~neiro Orioli, M.~Prüfer, and J.~Schmiedmayer for fruitful discussions. Significant parts of this work are taken from the master's thesis of T.V.Z at Heidelberg University. This publication is part of and supported by the DFG Collaborative Research Centre "SFB 1225 (ISOQUANT)".
\end{acknowledgments}

\appendix
\section{General Bogoliubov Transformation \label{appendix_bogoliubov}}
In this appendix we summarize the inhomogeneous Bogoliubov 
transformation for unstable systems (see also the appendix of \cite{garay2001sonic}). We consider a general $d-$dimensional Hamiltonian of the quadratic form
\begin{align}
H &= \int dx\, \phi^\dagger(t,x) H_0(x) \phi(t,x) \nonumber \\
&+ \frac{1}{2} \int dx\,\left(\phi^\dagger(t,x) H_1(x)\phi^\dagger(t,x) + \text{h.c.}\right)
\end{align}
with complex bosonic Heisenberg field operators satisfying canonical commutation relations
\begin{align}
\left[\phi(t,x), \phi^\dagger(t,y)\right] = \delta(x-y) 
\end{align}
and we require $H_0^\dagger = H_0$ for hermiticity.
The general Bogoliubov transformation reads
\begin{align}
\phi(t,x) = \sum_n \left(u_n(x) e^{-iE_n t} a_n + v_n^*(x) e^{iE_n^* t} a_n^\dagger\right) \; ,
\end{align}
where $a_n, a_n^\dagger$ are Schrödinger creation and annihilation operators spanning the Fock space in consideration. In general, they do not fulfill canonical commutation relations. Note also, that the eigenergies $E_n$ are in general complex numbers. To determine these, we consider the Heisenberg equations of motion in compact matrix form as
\begin{align}
\begin{pmatrix}
H_0 & H_1 \\ -H_1^\dagger & -H_0
\end{pmatrix}
\begin{pmatrix}
u_n \\ v_n
\end{pmatrix}
= E_n \begin{pmatrix}
u_n \\ v_n
\end{pmatrix} \; .
\end{align}
This is a non-hermitian eigenvalue problem for the doublet of mode functions $y_n = (u_n,v_n)$ known as the Bogoliubov-de Gennes equations. Using the hermitian conjugate of this equation we derive the general orthogonality relation between two eigenfunctions $y_n, y_m$ as
\begin{equation}
(E_n - E_m^*)\left(y_n,y_m\right) = 0 \; ,
\end{equation}
where we have defined the inner product
\begin{align}
\left(y_m,y_n\right) = \int dx \, \left(u_m^*(x) u_n(x) - v_m^*(x)  v_n(x)\right) \; .
\end{align}
By direct computation, we observe that for an eigenvector $y_n$ with eigenvalue $E_n$, there exist further eigenvectors with related eigenvalues
\begin{align}
y_n^* &= \begin{pmatrix} u_n^* \\ v_n^* \end{pmatrix} \; \text{with} \; E_n^* \; ,\\
\tilde{y}_n &= \begin{pmatrix} v_n \\ u_n \end{pmatrix} \; \text{with} \; -E_n \; ,\\
\tilde{y}_n^* &= \begin{pmatrix} v_n^* \\ u_n^* \end{pmatrix} \; \text{with} \; -E_n^* \; .
\end{align}
Labeling the corresponding quantum numbers as $n^*, \tilde{n}$ and $\tilde{n}^*$, respectively, we note that the Bogoliubov expansion contains terms like
\begin{align}
u_n(x) e^{-iE_n t} \left(a_n + a^\dagger_{\tilde{n}^*}\right) + v_n^*(x) e^{iE_n^* t} \left(a_n^\dagger + a_{\tilde{n}^*}\right)
\end{align}
showing a clear redundancy in the description. Redefining $a_n + a^\dagger_{\tilde{n}^*} \rightarrow a_n$ and dropping the corresponding $\tilde{\;}$ labels in the sum lifts this redundancy and we are left with two orthonormality relations
\begin{align}
\int dx \, \left(u_m(x) v_n(x) - v_m(x)  u_n(x)\right) &= 0 \; , \\
\int dx \, \left(u_m^*(x) u_n(x) - v_m^*(x)  v_n(x)\right) &= \delta_{nm^*}  \; ,
\end{align}
which hold for every eigenfunction $y_n(x) = \begin{pmatrix}u_n(x) \\ v_n(x)
\end{pmatrix}$ that solves the BdG equations. Now the eigenvalues appear in pairs $n, n^*$ with $E_{n^*} = E^*_n$. In the case of real $E_n$, we have $n = n^*$. We can use these relations to invert the Bogoliubov transformation as
\begin{align}
a_n(t) &= \int dx \; \left(u_{n^*}^* \phi - v_{n^*}^* \phi^\dagger\right)
\end{align}
and determine the commutation relations as
\begin{align}
\left[a_n,a_m^\dagger\right] = \delta_{nm^*} \; , && \left[a_n,a_m\right] = 0 \; .
\end{align}
We emphasize that these are not canonical commutation relations. In general, it is impossible to diagonalise an arbitrary quadratic Hamiltonian while retaining canonical commutation relations. Finally applying the Bogoliubov transformation, we can rewrite the Hamiltonian in diagonal form
\begin{equation}
H = \sum_n E_n a_n^\dagger a_{n^*} \; ,
\end{equation}
where we have dropped an irrelevant constant.
Note that the sum runs over all quantum numbers $n$ including their dual $n^*$ (for complex $E_n$).

\section{Simulation Details and Numerical Parameters \label{appendix_parameter}}
\begin{figure}
\includegraphics[scale = 0.15]{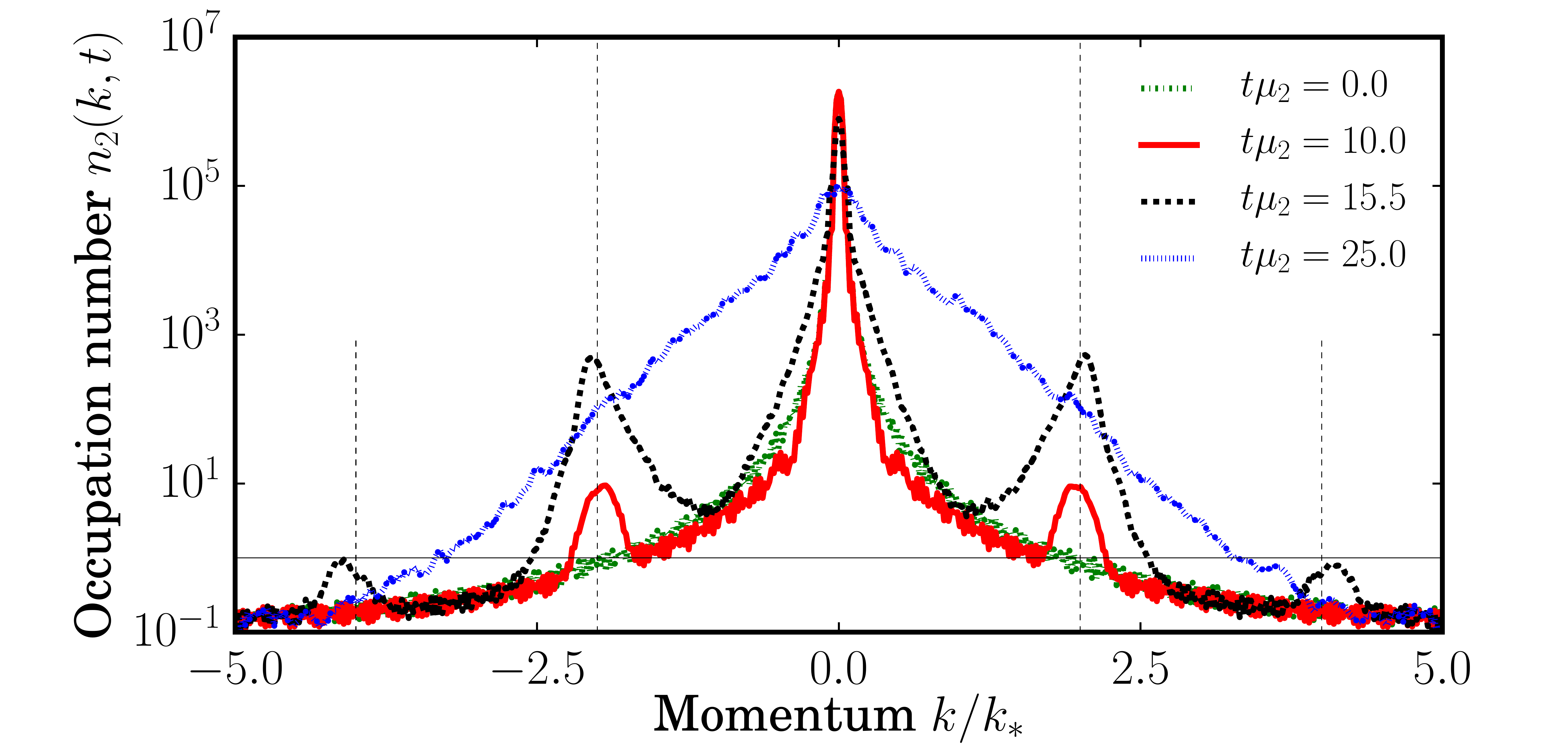}
\caption[Simulation 1: Secondary instabilities of the excited field]{\label{phi2_sim1}\emph{Simulation 1.} Spectrum of occupation of $\varphi_2$ at different times in units of $\bar{t} = \frac{\hbar}{\mu_2}$. The data has been smoothened with a Savitzky-Golay filter in order to reduce oscillations due to finite resolution. We clearly observe secondary instabilities at even multiples of the most unstable $k_*$ as indicated by the dashed vertical lines. The clear peak structure vanishes at late times and gives way to a broad occupation of many modes, which signals a non-perturbative regime (cf. figure \ref{field1_harmonic_unrealistic}).}
\end{figure}

\begin{figure}
\includegraphics[scale = 0.15]{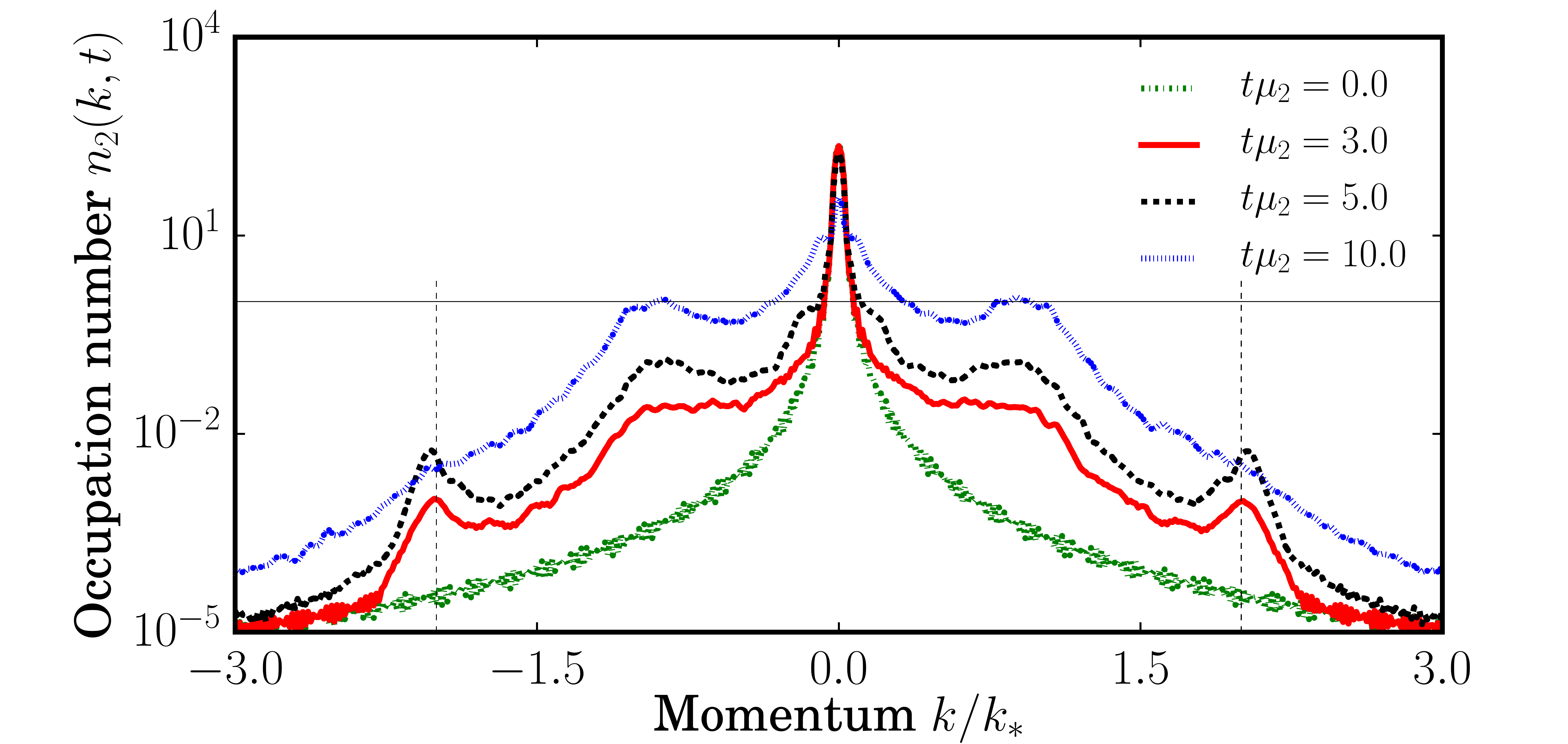}
\caption[Simulation 2: The excited field of the twin beam experiment]{\label{phi2_sim2}\emph{Simulation 2.} Spectrum of occupation of $\varphi_2$ at different times in units of $\bar{t} = \frac{\hbar}{\mu_2}$. The curves are smoothened like in the previous figure. As indicated by the horizontal line, no significant dynamics are detectable due to particle number limitations in the experimental setup. However, the numerical simulation reveals the possible excitation of secondaries at the expected $\pm 2 k_*$. The plateau-like structure with range $\left[-k_*,k_*\right]$ is an artifact of the truncated initial fluctuations in the lower field $\varphi_1$ (cf. figure \ref{realistic_sim}).}
\end{figure}

\begin{figure*}
\centering{
\includegraphics[width=0.45 \textwidth]{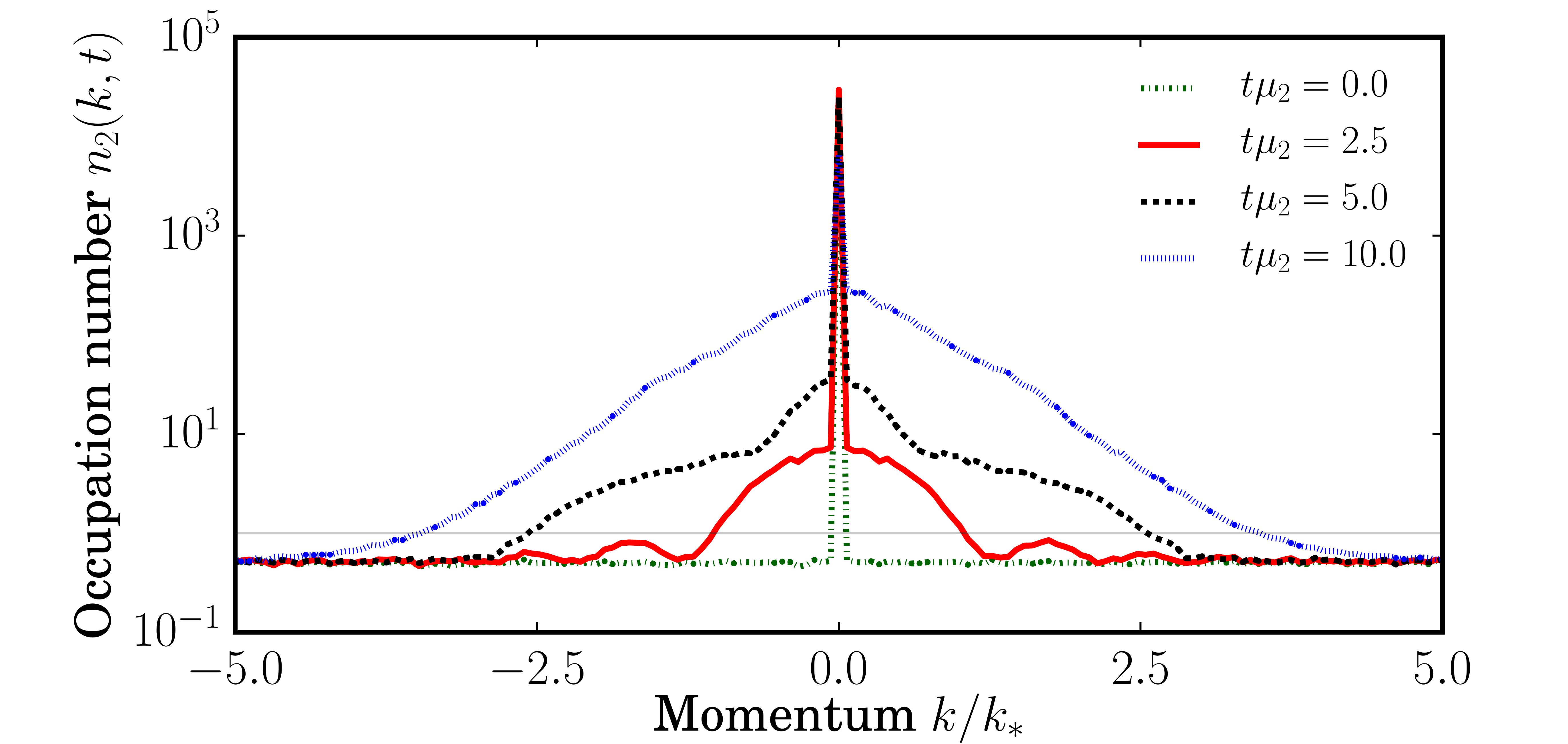}
\includegraphics[width=0.45 \textwidth]{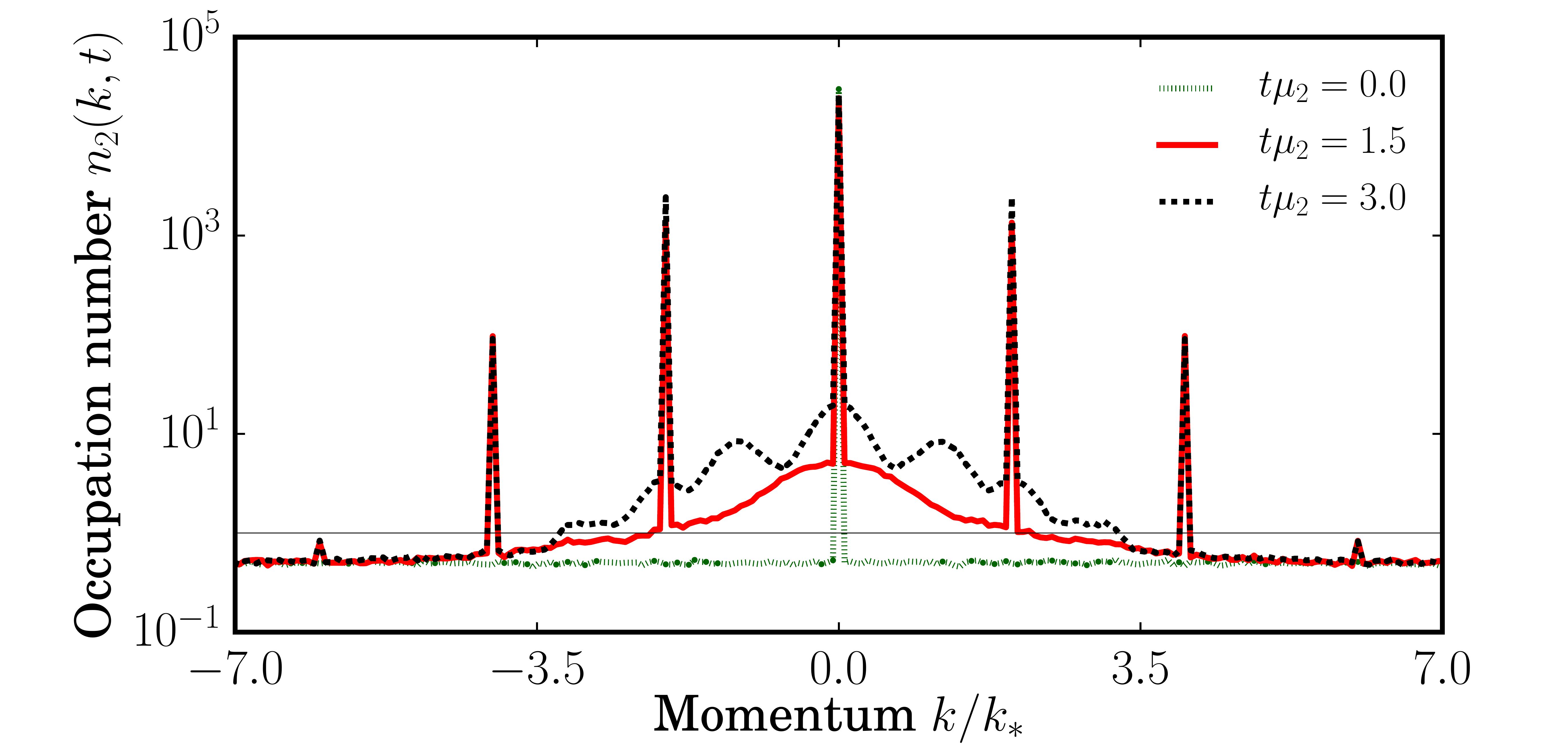}}
\caption[Simulation 3: Comparison of the excited field with and without seed]{\label{phi2_sim3}\emph{Simulation 3.} Comparison of the spectrum of occupation of $\varphi_2$ at different times in units of $\bar{t} = \frac{\hbar}{\mu_2}$ (cf. figure \ref{seed-noseed}). No smoothening of the curves was performed.\\ Left: No Seed. The dynamics occupy a broad range of momenta.\\
Right: Situation with seed. We observe the characteristic comb of peaks at even multiples of the most unstable $k_*$ indicating the cascade of secondaries as expected.}
\end{figure*}

For all simulations, we have considered Rubidium atoms with atomic mass $M = \SI{1.44e-25}{\kilogram}$ and scattering length $a_s = \SI{5.3}{\nano \meter}$. The interaction strength is given by $g = \frac{4 \pi \hbar^2 a_s}{M}$. We have used the classical-statistical or TWA approach to simulate the quantum dynamics numerically. For completeness, we also show the occupation of the excited field in figures \ref{phi2_sim1}, \ref{phi2_sim2} and \ref{phi2_sim3}.

\subsection{Proof of Principle: Generation of Secondaries}
The first simulation is based on the following parameters used to demonstrate the principle of secondaries instabilities. We used the particle number $N = \SI{6e6}{}$, the harmonic frequency $\omega = \SI{20}{\hertz}$ of the longitudinal trapping potential, $600$ lattice points with equal spacing $a= \SI{0.33}{\micro \meter}$ and assumed a cylindrical trap in orthogonal direction with radial size $R = \SI{2.7}{\micro \meter}$. We have reduced the atomic interaction as $\tilde{g} = g/500$. The two fields $\varphi_{1,2}$ arise as coefficients in an expansion of the 3D field into radial eigenfunctions as
\begin{align}
\psi(\mathbf{r}) &= \varphi_1(z)\psi_1(\rho, \phi) + \varphi_2 (z )\psi_2(\rho, \phi) + \dots
\end{align} 
where $\psi_1(\rho, \phi) = \psi_{0,1}(\rho, \phi)$ and $\psi_2(\rho, \phi) = \frac{1}{\sqrt{2}}\left[\psi_{1,1}(\rho, \phi) + \psi_{-1,1}(\rho, \phi)\right]$ are the energetically lowest lying states with $\psi_{l,n}$ denoting the single particle eigenfunctions of the cylindrical box. For simplicity, we implicitely assume a slight breaking of the cylindrical symmetry in order to lift the degeneracy of the first excited state. Then the interaction strengths are given by the following overlap integrals
\begin{subequations}
\begin{align}
g_1 &= \int_0^\infty d\rho \int_0^{2\pi} d\phi\, \rho \left|\psi_1(\rho,\phi)\right|^4 \nonumber\\
&= 1.049 \times \frac{2\tilde{g}}{\pi R^2} \; , \\
g_2 &= \int_0^\infty d\rho \int_0^{2\pi} d\phi\, \rho \left|\psi_2(\rho,\phi)\right|^4 \nonumber\\
&= 0.7758 \times \frac{3\tilde{g}}{\pi R^2}  \; ,\\
g_{12} &= \int_0^\infty d\rho \int_0^{2\pi} d\phi\, \rho \left|\psi_1(\rho,\phi)\right|^2 \left|\psi_2(\rho,\phi)\right|^2 \nonumber\\
&= 0.7176 \times \frac{2\tilde{g}}{\pi R^2} \;.
\end{align}
\end{subequations}
The chemical potentials are fixed by the total particle number and the energy difference between the one-particle states in the cylindrical potential, i.e.
\begin{subequations}
\begin{align}
\mu_2 &= \left(\frac{3}{4} \tilde{g} N \sqrt{\frac{m}{2}\omega^2}  \right)^{3/2} \; ,\\
\mu_1 &= \mu_2 + 24.69 \frac{\hbar^2}{2m R^2}\,.
\end{align}
\end{subequations}
We have seeded the lower state $\varphi_1$ with quantum fluctuations and used the first 150 eigenfunctions of the harmonic oscillator as a controlled UV-cutoff. One simulation consists of 50 samples of initial values and their time evolution. We employed fixed boundary conditions and used a total simulation time of $T = 25 \frac{\hbar}{\mu_2}$ and readout steps of $dt = 0.1 \frac{\hbar}{\mu_2}$.

\subsection{Analysis of the real Twin-Beam Experiment}
The comparision to the twin-beam experiment was done with the following parameters: $N = 800$ particles, harmonic frequency $\omega = 2\pi \times \SI{16.3}{\hertz}$ and harmonic radial traps with $\omega_x = 2\pi \times \SI{1.83}{\kilo \hertz}, \omega_y = 2\pi \times \SI{2.5}{\kilo \hertz}$, giving the oscillator lengths $a_i = \sqrt{\frac{\hbar}{m \omega_i}}$. We used $600$ lattice points with equal spacing $a = \SI{0.15}{\micro \meter}$. This time the fields $\varphi_{1,2}$ are defined via
\begin{align}
\psi(\mathbf{r}) &= \varphi_1(z)\psi_0(x)\psi_0(y) + \varphi_2 (z )\psi_0(x)\psi_1(y) + \dots
\end{align}
with $\psi_n$ denoting harmonic oscillator eigenfunctions which results in the following interaction constants
\begin{subequations}
\begin{align}
g_1 &= \int dx \int dy |\psi_0(x)|^4 |\psi_0(y)|^4 = \frac{g}{2\pi a_x a_y} \; ,\\
g_2 &= \int dx \int dy |\psi_0(x)|^4 |\psi_1(y)|^4 =\frac{3g}{8\pi a_x a_y}\; ,\\
g_{12} &= \int dx \int dy |\psi_0(x)|^4 |\psi_0(y)|^2 |\psi_1(y)|^2 = \frac{g}{4\pi a_x a_y} \, .
\end{align}
\end{subequations}
The chemical potentials are fixed as $\mu_2 = \SI{492}{\hertz} \times h$ and $\mu_1 = \mu_2 + \hbar \omega_x$. The total simulation time is $T = 10 \frac{\hbar}{\mu_2}$ in steps of $dt = 0.1 \frac{\hbar}{\mu_2}$. We used the first $151$ eigenfunctions to seed the initial fluctuations in $\varphi_1$ for 100 samples and employed fixed boundary conditions.

\subsection{Proposal of Seeding Procedure}
For the direct comparison of the situation with and without seed, we used the following parameter set: $N_2 = 30000$ particles in $\varphi_2$, a system length of $L = \SI{80}{\micro \meter}$ (no longitudinal trap!) and a cylindrical trap size of $R = \SI{3}{\micro \meter}$. The same reasoning as for the first simulation gives the effective interaction constants $g_1, g_2, g_{12}$ from overlap integrals (without tilde, i.e. we used the true $g$ instead of $\tilde{g}$). We used $300$ lattice sites with equal spacing $a =\SI{0.13}{\micro \meter}$. The chemical potentials read $\mu_2 = g_2 N / L$ and $\mu_1$ is fixed as in the first simulation. The total simulation time is $T = 10 \frac{\hbar}{\mu_2}$ in steps of $dt = 0.05 \frac{\hbar}{mu_2}$. We employed periodic boundary conditions, which is justified for sufficiently large homogeneous systems. One simulation consists of 100 samples of initial fluctuations.\\
Without seed: Only initial fluctuations in $\varphi_1$.\\
With seed: Fluctuations on top of seed modulation at $k_* \approx 15 \times \frac{2 \pi}{a}$ in $\varphi_1$ with fraction $\beta = \frac{1}{2}$ of $\varphi_2$.

\subsection{Classical Toy Model}
The simple four-mode model featuring secondaries derived from the coupled GP equations is given by
\begin{align}
i \partial_t \varphi_j = \mathcal{H}_j\left[\varphi_0, \varphi_1, \varphi_2, \varphi_3\right]
\end{align}
with $\mathcal{H}_j\left[\varphi_0, \varphi_1, \varphi_2, \varphi_3\right] = \mathcal{H}_{\text{lin}}^{(j)}[\varphi_j] + \mathcal{H}^{(j)}_\text{NL}\left[\varphi_0, \varphi_1, \varphi_2, \varphi_3\right]$. The linear and nonlinear terms are given by
\begin{subequations}
\begin{align}
\mathcal{H}_\text{lin}^{(0)} [\varphi_0] &= -\mu_2 \varphi_0 \; ,\\
\mathcal{H}_\text{lin}^{(1)} [\varphi_1] &= -2 \gamma_* \varphi_1 \; ,\\
\mathcal{H}_\text{lin}^{(2)} [\varphi_2] &=   \left(\frac{2k_*^2}{M} - \mu_2\right)\varphi_2 \; ,\\
\mathcal{H}_\text{lin}^{(3)} [\varphi_3] &= \left(\frac{4 k_*^2}{M} - 2\gamma_*\right) \varphi_3 
\end{align}
\end{subequations}
and
\begin{subequations}
\begin{align}
\mathcal{H}_\text{NL}^{(0)} &= c_2 \left\lbrace\varphi_0^*\varphi_0^2 + \varphi_2^*\left[4\varphi_0\varphi_2 + 2\varphi_2^2\right] \right\rbrace \nonumber\\ &+ c_{12} \left\lbrace \varphi_0^*\left[2\varphi_1^2 + 2 \varphi_3^2\right] + \varphi_2^* \left[2\varphi_1^2 + 4 \varphi_1\varphi_3\right] \right\rbrace  \nonumber\\ &+ 2c_{12} \left\lbrace \varphi_1^* \left[2\varphi_1\varphi_0 + 2\varphi_1\varphi_2 + 2\varphi_3\varphi_2\right] \right\rbrace \nonumber\\ &+ 2c_{12}\left \lbrace \varphi_3^* \left[2\varphi_1\varphi_2 + 2\varphi_3 \varphi_0\right] \right\rbrace   \; ,\\
\mathcal{H}_\text{NL}^{(1)} &=  c_1 \left\lbrace \varphi_1^*\left[3\varphi_1^2 + 2\varphi_1\varphi_3 + 2\varphi_3^2\right] + \varphi_3^* \left[2\varphi_1^2 +4\varphi_1 \varphi_3\right] \right\rbrace \nonumber\\ &+ c_{12}\left\lbrace \varphi_1^*\left[\varphi_0^2 + 2\varphi_0 \varphi_2 + 2 \varphi_2^2\right] + \varphi_3^*\left[2\varphi_0 \varphi_2 + \varphi_2^2\right]\right\rbrace \nonumber\\ &+ 2c_{12} \left\lbrace \varphi_0^*\left[\varphi_0\varphi_1 + \varphi_2\varphi_1 + \varphi_2\varphi_3\right] \right\rbrace \nonumber\\ &+ 2c_{12} \left\lbrace \varphi_2^* \left[\varphi_0\varphi_1 + \varphi_0\varphi_3 + 2\varphi_2\varphi_1 + \varphi_2\varphi_3\right]\right\rbrace \; ,\\
\mathcal{H}_\text{NL}^{(2)}  &=  c_2 \left\lbrace \varphi_0^* \left[2\varphi_0\varphi_2\right] +\varphi_2^*\left[\varphi_0^2 + 3\varphi_2^2\right] \right\rbrace  \nonumber\\ &+ c_{12}\left\lbrace \varphi_0^* \left[\varphi_1^2 + 2\varphi_1\varphi_3\right] +\varphi_2^* \left[2\varphi_1^2 + 2\varphi_1\varphi_3 + 2\varphi_3^2\right]\right\rbrace  \nonumber\\ &+ 2c_{12}\left\lbrace \varphi_1^*\left[\varphi_1\varphi_2 + \varphi_1\varphi_0 + \varphi_3\varphi_2 + \varphi_3\varphi_0\right] \right\rbrace  \nonumber\\ &+ 2c_{12}\left\lbrace \varphi_3^* \left[\varphi_1\varphi_0 + \varphi_1\varphi_2 + 3\varphi_3\varphi_2\right]  \right\rbrace   \; ,\\
\mathcal{H}_\text{NL}^{(3)}  &= c_1 \left\lbrace \varphi_1^* \left[\varphi_1^2 + 4\varphi_1 \varphi_3\right] + \varphi_3^* \left[2\varphi_1^2 + 3\varphi_3^2\right] \right\rbrace \nonumber\\ &+ c_{12} \left\lbrace \varphi_1^* \left[2\varphi_0\varphi_2 + \varphi_2^2\right] +\varphi_3^* \left[\varphi_0^2 +2\varphi_2^2\right]\right\rbrace \nonumber\\ &+ 2c_{12} \left\lbrace \varphi_0^* \left[\varphi_0\varphi_3 + \varphi_2 \varphi_1\right] \right\rbrace \nonumber\\ &+ 2c_{12} \left\lbrace\varphi_2^* \left[\varphi_0\varphi_1 +\varphi_2\varphi_1 + 2\varphi_2\varphi_3\right] \right\rbrace   \; .
\end{align}
\end{subequations}

For the simulation we used units of $\mu_2$, which sets the physical time scale $\bar{t} = \frac{\hbar}{\mu_2}$. We consider a weak dimensionless coupling $\frac{c_2}{\mu_2} = \SI{1e-10}{}$ and $\frac{c_1}{c_2} = 11$, $\frac{c_{12}}{c_2} = 7$ which are related to physical couplings via $g_X = c_X L$ with a ficticious system length $L$ and $X = 1, 2, 12$. Note that the growth rate $\gamma_*$ is independent of $L$. Furthermore, we chose $\frac{k_*^2}{2M\mu_2} = 3$. The initial values of the non-condensate modes are all equal and suppressed with respect to the condensate mode by $\beta = \SI{e-5}{}$.

\section{Quantum Evolution Equations \label{app:evol_eqs}}
Rephrasing the two-field system of \eqref{eq:Hamiltonian1} in terms of effective action techniques, we consider the corresponding classical action
\begin{align}
S[\phi] = \int_{\mathcal{C}}dt \; \left\lbrace\sum_{a=1,2} \int dz \; i\phi_a^* \partial_t  \phi_a -H[\phi] \right\rbrace \; ,
\end{align}
where $\phi$ denotes classical fields in contrast to the Heisenberg operators $\varphi$. The dynamics of the quantum theory are encoded in the generating functional $Z[J, R] = \int \mathcal{D}\phi \; e^{i\left(S[\phi]+ J \cdot \phi + \phi \cdot R \cdot \phi \right)}$ and $\mathcal{C}$ is the Schwinger-Keldysh contour. We employ the 2PI effective action $\Gamma\left[\bar{\phi},G\right]$, which is defined as the Legendre transform of $W[J,R] = -i \log Z[J,R]$ w.r.t. the sources $J,R$. In this formulation the quantum equations of motion for the macroscopic field $\bar{\phi}$ and the full propagator $G$ are given by
\begin{align}
\frac{\delta \Gamma}{\delta \bar{\phi}} = 0 \; , && \frac{\delta \Gamma}{\delta G} = 0 \; . 
\end{align}
In addition to the macroscopic field
\begin{align}
\bar{\phi}_\alpha(t,z) &= \langle \varphi_\alpha(t,z) \rangle \; ,
\end{align}
it is convenient to introduce the \emph{statistical propagator} $F$ and the \emph{spectral function} $\rho$ as
\begin{align}
F_{\alpha \beta}(t,z;t',z') &= \frac{1}{2} \left\langle \lbrace \varphi_\alpha(t,z), \varphi_\beta(t',z')\rbrace \right\rangle \nonumber\\
&\quad - \bar{\phi}_\alpha(t,z)\bar{\phi}_\beta(t',z') \; , \\
\rho_{\alpha \beta}(t,z;t',z') &= i \left\langle \left[ \varphi_\alpha(t,z), \varphi_\beta(t',z')\right] \right\rangle \; ,
\end{align}
which are the symmetric $(F)$ resp. anti-symmetric $(\rho)$ parts of $G$ w.r.t. $\mathcal{C}$. Note that here and in the following $\alpha, \beta = 1,2,3,4$ label $\varphi_1, \varphi_1^\dagger, \varphi_2, \varphi_2^\dagger$. In this basis, one can rewrite the quantum evolution equations as
\begin{widetext}
\begin{align}
\mathcal{D}_{\alpha \gamma}(t,z) F_{\gamma \beta}(t,z;t',z') &= - \int_{t_0}^{t} ds \int dy \; \Sigma_{\alpha \gamma}^\rho(t,z;s,y) F_{\gamma \beta}(s,y;t',z') + \int_{t_0}^{t'} ds \int dy \; \Sigma_{\alpha\gamma}^F(t,z;s,y) \rho_{\gamma \beta}(s,y;t',z')\\
\mathcal{D}_{\alpha \gamma}(t,z) \rho_{\gamma \beta}(t,z;t',z') &= -\int_{t'}^{t} ds \int dy \; \Sigma_{\alpha \gamma}^\rho (t,z;s,y) \rho_{\gamma \beta}(s,y;t',z')
\end{align}
\end{widetext}
supplemented by an evolution equation for the macroscopic field, which is not relevant for the present dicussion.
The differential operator $\mathcal{D}$ has a contribution from the inverse classical propagator
\begin{align}
iG_{0,\alpha \beta}^{-1}(t,z;t',z';\bar{\phi}) &= \frac{\delta^2 S[\bar{\phi}]}{\delta \bar{\phi}_\alpha(t,z)\delta \bar{\phi}_\beta(t',z')} \nonumber\\
&= -D_{0,\alpha \beta}(t,z) \delta(t-t')\delta(z-z') \; ,
\end{align}
where in $4 \times 4$-matrix notation the operator
\begin{align}
D_0(t,z; \bar{\phi}) = \begin{pmatrix}
\sigma^1 H_0^{(1)} + \sigma^2 \partial_t + I_{1} & \sigma^1 H_{12} + I_{12}\\
\sigma^1 H_{12} + I_{12} & \sigma^1 H_0^{(2)} + \sigma^2 \partial_t +I_{2}
\end{pmatrix}
\end{align}
and $\sigma^j$ denote Pauli matrices and we have abbreviated
\begin{subequations}
\begin{align}
H_0^{(1)}[\phi] &= H_0 + 2g_1 |\phi_1|^2 +2 g_{12}|\phi_2|^2 - \mu_1 \;  ,\\
H_0^{(2)}[\phi] &= H_0 + 2g_2 |\phi_2|^2 +2 g_{12}|\phi_1|^2 - \mu_2 \;  ,\\
H_{12}[\phi] &= 2g_{12}\left(\phi_2^* \phi_1 + \phi_1^* \phi_2\right) \;  ,\\
H_0 &= -\frac{\Delta_z}{2M}\;  ,\\
I_{1}[\phi] &= \begin{pmatrix}
g_{12}\left(\phi_2^*\right)^2 + g_1 \left(\phi_1^*\right)^2 & 0\\
0 & g_{12} \phi_2^2 + g_1 \phi_1^2
\end{pmatrix}\;  ,\\
I_{2}[\phi] &= \begin{pmatrix}
g_{12}\left(\phi_1^*\right)^2 + g_2 \left(\phi_2^*\right)^2 & 0\\
0 & g_{12} \phi_1^2 + g_2 \phi_2^2
\end{pmatrix}\;  ,\\
I_{12}[\phi] &= \begin{pmatrix}
2g_{12} \phi_1^* \phi_2^* &0 \\ 0& 2g_{12}\phi_1 \phi_2
\end{pmatrix}
\end{align}
\end{subequations}
and the operator
\begin{align}
\mathcal{D}_{\alpha \beta}(t,z) = -iD_{0,\alpha \beta}(t,z) + \Sigma_{\alpha \beta}^{(0)}(t,z) \; .
\end{align}
As such, the equations \eqref{2pi1} \& \eqref{2pi2} are exact. In order to make progress one has to specfiy an approximation for the self-energies $\Sigma$, which are split into local $\Sigma^{(0)}$ and non-local parts $\Sigma^{\rho/F}$. The superscripts $F$ resp. $\rho$ denote symmetric resp. anti-symmetric contributions to $\Sigma$ w.r.t. the closed time-contour.

Setting $\Sigma = 0$ (tree-level) and fixing the macroscopic fields $\bar{\phi}_2 = \sqrt{\frac{\mu_2}{g_2}}$, $\bar{\phi}_1 = 0$, we find
\begin{align}\label{tree-level prop}
i \partial_t \begin{pmatrix}
F_{11}(t,z;t',z') \\ F_{12}(t,z;t',z')
\end{pmatrix} = \begin{pmatrix}
H' & \gamma_*\\
-\gamma_* & -H'
\end{pmatrix}\begin{pmatrix}
F_{11}(t,z;t',z') \\ F_{12}(t,z;t',z')
\end{pmatrix} \; ,
\end{align}
where $H' = H_0 + 2g_{12} \frac{\mu_2}{g_2} - \mu_1$
and similiar for $F_{21}, F_{22}$ from which we recover the Bogoliubov result \eqref{disp_rel}. As long as $\bar{\phi}_1 = 0$, the $1,2$ components decouple from the $3,4$ components of $F$ and e.g. $F_{31} \equiv 0$. We can derive a similar equation,
\begin{align}
i \partial_t \begin{pmatrix}
F_{33} \\ F_{34}
\end{pmatrix} = \begin{pmatrix}
H_0 + \mu_2 & \mu_2\\
-\mu_2 & -\left(H_0  + \mu_2\right)
\end{pmatrix}\begin{pmatrix}
F_{33} \\ F_{34}
\end{pmatrix} \; ,
\end{align}
from which one can read off the dispersion relation for the excited field as
\begin{align}
\tilde{\omega}_k^2 = \left(\frac{k^2}{2M}\right)^2 + 2\mu_2 \frac{k^2}{2M} >0 \; ,
\end{align}
which is manifestly stable.

\newpage

\bibliographystyle{apsrev4-1}
\bibliography{0publications.bib}

\end{document}